\documentclass[]{interact}
\usepackage{subfigure}
\usepackage{amsmath}
\usepackage{graphicx}
\usepackage{dcolumn}
\usepackage{bm}
\usepackage{soul,color}
\usepackage{times}
\usepackage[numbers,sort&compress,merge]{natbib}

\theoremstyle{plain}

\theoremstyle{definition}

\theoremstyle{remark}

\begin{document}

\title{Weak measurement effect on optimal estimation with lower and upper bound on relativistic metrology }

\author{
\name{H. Rangani Jahromi \thanks{ Email: h.ranganijahromi@jahromu.ac.ir}}
\affil{Physics Department, Faculty of Sciences, Jahrom
University, P.B. 7413188941,  Jahrom, Iran.}
}

\maketitle

\begin{abstract}
We address the quantum estimation of parameters encoded into the initial state of  two modes of a Dirac field described by relatively accelerated
parties. By using the quantum Fisher information (QFI),  we investigate how  the weak measurements performed before and after the accelerating observer,  affect the optimal estimation of information encoded into the weight and phase parameters  of the initial state shared between the parties.  Studying the QFI, associated with  weight parameter $ \vartheta $, we find that  the acceleration at which the optimal estimation occurs may be controlled by weak measurements. Moreover, it is shown that the post-measurement plays the role of a quantum key for manifestation of  the Unruh effect. On the other hand, investigating the phase estimation optimization and assuming that there is no control over  the initial state, we show that the weak measurements may be utilized to match the optimal $ \vartheta $ to  its predetermined  value. Moreover, in addition to determination of a lower bound on the QFI with the local quantum uncertainty,  we  unveil an important upper bound on the precision of phase estimation  in our relativistic scenario, given  by the maximal steered coherence (MSC).  We also obtain a  compact expression of the MSC for general X states.
\end{abstract}


\section{Introduction\label{introduction}}
Quantum metrology, investigating the estimation of quantities not corresponding to observables of the given quantum system, has attracted a great deal of attention recently \cite{Helstrom1976,M.G.A. Paris,Xiao-Ming Lu,Lina Chang,Zhang Jiang,Wei Zhong,J. Ma,Z. Sun,Yao Yao,V. GiovannettiPRL,RanganiAOP,Jian Ma,RanganiOPTC,RanganiAOP2}.
It intends to yield higher
statistical precision of an unknown parameter by using and controlling
 quantum resources rather than those resources available only
 in classical approaches  \cite{Giovannetti2011}. There have been
many studies of
parameter estimation  in different physical systems,  such as optical interferometry \cite{Israel2014,Xiang2011}, atomic
systems \cite{RanganiAOP,Schleier2011}, and Bose-Einstein condensates \cite{Sorensen2001}. 
The QFI, playing a central role in quantum metrology and information
theory,  quantifies the maximum precision  achievable by picking the optimal estimation strategy. In fact, the QFI determines the
lower bound on the variance of an unbiased estimator for the parameter under estimation,
according to the quantum Cram\'{e}r-Rao  theorem \cite{Helstrom1976}. Moreover, QFI has different applications in quantum technology like clock synchronization \cite{R. Joza2000},  quantum frequency standards \cite{Bollinger1996}, and gravity acceleration measurement \cite{Peters400}. The standard estimation process, following in this paper, involves these three steps: (I)  encoding the unknown parameters into a two-qubit system; (II) interaction of the system with the environment; (III) final measurement on the system to extract the encoded information.
One  looks for the most optimal measurement strategy achievable in practice such that the error in the estimation process is minimized.

\par
 As a novel compound, composed of information
theory, quantum field theory, and general relativity,  relativistic quantum information \cite{Fuentes2005,Adesso2007,Leon2009,Adesso2009,EMM0062010,EMM3182009,EMM3202010,Fuentes0302010,EMM0282010,EMM3052010,Bruschi3322010,EMM2010,Alsing2012,EMM5012013,Lee0412014,Lanzagorta2014,Friis222012} not only is remarkably
important in studies of long-distance quantum communication placed in curved background spacetimes, but also reveals  new aspects of  close and complex relationship
between general relativity and quantum mechanics. Hence,  it is interesting to discuss  QFI in the relativistic framework \cite{Aspachs2010,Downes2011,Hosler2013,Ahmadi9962014,Ahmadi0282014,Yao2014,Wang2014,Sabin2014,Doukas2014,Bruschi0012014,Safranek2015,Kish2016,Huang2017,Huang2018,RanganiQIP4,Kohlrus2019}. 
For example, in Ref. \cite{Yao2014}, the authors investigated the QFI of two-qubit systems for
both Dirac and scalar fields when one observer is accelerated and found that for both cases, the QFI with respect
to different state parameters exhibits diverse properties.
In detail, they show that the QFI with respect to the phase parameter  shows a decay behaviour with increasing the acceleration, while the QFI with respect to the weight parameter is completely unaffected by the acceleration.
As another example,  Huang \textit{et al}. \cite{Huang2017}, focus on exploring the
bahaviour of QFI and non-locality of Dirac particles in multipartite systems in which more
than one observer are accelerated.
They mainly investigate the difference between
QFI and non-locality of a multipartite state in non-inertial frame. Moreover,  the quantum metrology for a pair of entangled Unruh-Dewitt detectors when one of them is
accelerated and coupled to a massless scalar field has been studied in \cite{Wang2014}. Recently,  partial (weak) measurements \cite{kim2012} have been proposed as means to achieve  enhancements
in quantum metrology  for some physical systems \cite{Chen2017,Zhi2017}. However, the effects of partial measurements on the quantum parameter estimation in the relativistic scenario has not been investigated in detail yet. This is the line that will be followed in our paper by  investigating the effects weak measurements on the Unruh effect, causing a uniformly accelerated detector, interacting with external fields, to become excited in the Minkowski vacuum \cite{Wang2014}.

\par
In quantum information theory,
nonclassical correlations, usually quantified by  quantum discord \cite{Modi2012} may 
plays a key role   in quantum metrology.
The computation of 
discord for a general two-qubit state is very difficult and compact analytical expressions have been found
only for some special types of states.
Nevertheless, a discord-like measure of quantum correlation,
precisely computable for any two-qubit  system, i.e., local quantum uncertainty
(LQU), has been proposed   recently \cite{Girolami2013,Lecture}.
Not only does the LQU  measure the quantum
correlation, but also it may apply to the field of quantum metrology \cite{Girolami2013,Yu2014}. In Ref. \cite{Girolami2013}, the relation between the QFI and LQU has been discussed in the unitary evolution. In this scenario, the amount of discord (LQU) in a mixed correlated  state $ \rho $, used for estimating parameter $ \varphi $,
bounds from below the squared speed of evolution of the
state under any local Hamiltonian evolution $e^{-i\varphi H_{A}}  $. On the other hand,   a higher
speed of state evolution under a change in parameter $\varphi $,
corresponds to a higher sensitivity of the given probe state to the
parameter estimation. Moreover, Ref. \cite{Shao2018} has been studied the relationship between the QFI and the LQU in a special
open quantum system involving two coupled qubits interacting with the
independent non-Markovian Lorentzian form environments.

Another important property which its relationship with  parameter estimation should be investigated, is quantum coherence originating from the quantum pure state superposition principle. It is  recognized as a significant resource in some scenarios,
including quantum reference frames \cite{Bartlett2007}, quantum thermodynamics \cite{Narasimhacha2014}, and transport in  biological systems \cite{Lloyd2011}.  In particular,
 supposing that Alice and Rob share a correlated  bipartite quantum state, the authors of \cite{Hu2016} investigated the coherence of 
Rob's \textit{steered} state, which is obtained by Alice's measurement. Because the steering directly originates from the quantum correlations whose effects on parameter estimation have been widely studied \cite{RanganiQIP3,R. Demkowicz,Rangani64}, we are motivated to explore the relation between QFI and coherence of the steered state.

Partial measurements \cite{Korotkov2006},  generalizations of the usual  von
Neumann measurements, keep the measured state alive, because it does not completely collapse towards an eigenstate.
Therefore, it is possible to retrieve the initial information, encoded into the quantum state of the system,
with some operations, even when the quantum state has been affected by decoherence. 
Recently, many proposals, investigating
the partial measurements for protecting the   a single qubit fidelity, the
quantum entanglement of two qubits, and two qutrits from amplitude damping  decoherence have been demonstrated
both theoretically and experimentally  \cite{Sun2009,Man2012,XiaoEuro2013,Neeley2008,KimExpress,kim2012}. Moreover, in Ref. \cite{XiaoPRA2016}, enhancement of  QFI teleportation by partial measurements has been studied. Besides, improvement of the QFI transmission via quantum channel composed of
the spin chain consisting of  interacting spin-1/2 particles by partial measurements has been discussed in \cite{ZhiLiu2017}. In addition,  a scheme has been proposed  by using partial measurements
to protect the average QFI in the independent amplitude-damping
channel  for $ N $-qubit GHZ states \cite{YuChen2017}. This
motivates us to study the parameter estimation in the presence of the Unruh effect 
by utilizing the partial measurements. On the other hand, it is worthy to
note that the relativistic causality would be violated when the projective measurements are performed between the causally separated
modes \cite{Benincasa2014}.  Although it has been  found that the violation can be
suppressed by introducing restrictions on the post-measurements for
the projective measurements on relativistic nonlocal modes \cite{Lin2013}, the problem has not been generally solved yet. Accordingly, investigating the role of weak measurements in realizing relativistic quantum information tasks (such as relativistic quantum metrology) deserves much more attention. 

\par
According to the classical
parameter estimation theory, it is well known that for a series of $M $ independent measurements of a random variable, the
minimum mean square error scales like $ 1/M $ with a proportionality coefficient  equal to the inverse of the Fisher
information (FI) \cite{Helstrom1976}. Therefore, the estimation problem
reduces   to selecting the measurement which promises the lowest estimation error, as encoded by the corresponding FI which should be maximized in the quantum theory over all
possible measurement strategies, obtaining the QFI. In fact, the QFI can be defined as the
upper bound of the FI corresponding to any
possible measurement provided that the measurement strategy aimed
at estimating the parameter does not depend on its value. However, there are many estimation scenarios in which the above condition does not hold and an alternative approach should be
developed in order to find a proper bound to the ultimate precision allowed by quantum mechanics \cite{Seveso2017}.

\par In this paper we address   the effects of partial measurements on quantum metrology of a two-qubit system in  relativistic quantum field theory. In particular, the  enhancement  of the parameter estimation  by partial measurements, carried out before and after appearance of the Unruh effect   is investigated. Moreover, the optimal behaviour of the QFI is studied analytically and a criterion for existence of the QFI optimal value is proposed. It is also explored how we can vary the points at which the optimal value of the QFI  is obtained by partial measurements. We also investigate the 
 lower bound on the QFI with the local quantum uncertainty in the presence of the Unruh effect. Moreover,  for the first time, we  unveil an upper bound on the precision of phase estimation, given  by the maximal steered coherence.  In addition, an important  compact
   expression of  this quantity for general X states is obtained.

  \par This paper is organized as follows: In Sec. \ref{Preliminaries} we give a brief description of
  LQU, quantum steering, QFI, and weak measurements. The physical model is presented in Sec. \ref{Model}.
  We study the effects of weak measurements on the optimal behaviour of the QFI in Sec. \ref{PE}. Moreover, in this section the lower and upper bound on the QFI  are obtained. Sec. \ref{conclusion} is devoted to
  conclusion. Besides,  analytical
  expression of maximal steered coherence for general X-states is obtained in Appendix \ref{A1}.

\section{The Preliminaries \label{Preliminaries}}

\subsection{Local Quantum Uncertainty}

By focusing on finite dimensional quantum systems,
let us suppose that we intend  to measure the observable being represented by a 
Hermitian operator $ O $ when the system is prepared in the state corresponding to density matrix $ \rho $.  If the
state is an eigenstate or a mixture of eigenstates of the observable, the operators $\rho $ and $ O $ commute \cite{Bogaert} and
hence  there is no change in the state after measurement, provided that  we focus on the von Neumann measurement model. 
 Under this condition  observable 
 $ O $ is dubbed \textit{quantum certain}. It is shown  that not only
entangled states but also almost all (mixed) separable
states  cannot admit any quantum-certain local observable. 

\par
One of the  ways to quantify the uncertainty when performing a 
measurement is the variance. However, the variance and entropic uncertainty quantifiers  include a contribution of classical uncertainty, for mixed states. It is easy to see that the variance may not vanish even if $ \rho $ and $ O $ commute, while a good measure of quantum uncertainty 
should be zero if and only if they commute \cite{Lecture}. 
It has been proposed that, extracting the truly
quantum share in quantifying the uncertainty on a
measurement, we can reliably quantify it via the \textit{skew information} \cite{Girolami2013}:
          \begin{equation}\label{skew}
I(\rho,O)=-\dfrac{1}{2} \text{Tr}\{[\sqrt{\rho},O]^{2}\}.
          \end{equation}
The skew information is upper bounded by the variance $V(\rho,O)=\text{Tr}\big[\rho O^2\big]-(\text{Tr}\big[\rho O\big])^2  $, being equal to it for pure states.  As an important concept in this analysis,  the
LQU is introduced as the minimum skew
information achievable on a single local measurement. We remind  that by \textit{measurement} in this section, we 
refer to a complete von Neumann measurement. Focusing on a bipartite system prepared in the state $ \rho=\rho_{AB} $, we  suppose that $ O^{\varLambda}\equiv O^{\varLambda}_{A} \otimes \mathcal{I} _{B}  $ represents a local observable, where $  O^{\varLambda}_{A}$ denotes a Hermitian operator on
$ A $ with nondegenerate spectrum $ \varLambda $.

 The LQU with respect to subsystem $ A $, optimized
over all local observables of $ A $ with nondegenerate spectrum $ \varLambda $, is defined by \cite{Girolami2013}

          \begin{equation}
\mathcal{U}^{\varLambda}_{A}=\min_{O^{\varLambda}}I(\rho,O^{\varLambda}).
          \end{equation}

Restricting ourselves to the case where  subsystem A is a qubit and B a qudit, we can find  that the choice
of the spectrum $ \varLambda $ does not affect the quantification of non-classical correlations, and hence 
we shall drop the $ \varLambda$ superscript from here onwards. Moreover, for qubit-qudit systems, we can write the LQU in the following form:

          \begin{equation}
\mathcal{U}_{A}=1-\lambda_{max}\big(W_{AB}\big),
          \end{equation}

where $\lambda_{max}\big(W_{AB}\big)  $  the maximum eigenvalue of the $ 3\times 3 $ symmetric matrix $ W $ with
elements:

          \begin{equation}
(W_{AB})_{ij}=\text{Tr}\big[\sqrt{\rho_{AB}}(\sigma_{iA}\otimes \mathcal{I}_{B}) \sqrt{\rho_{AB}}(\sigma_{jA}\otimes \mathcal{I}_{B})\big ],
          \end{equation}
 where i, j label the Pauli matrices.

\subsection{Quantum Steering Elipsoid}
The quantum steering ellipsoid of a two-qubit state is the set of Bloch vectors that Alice can collapse
Rob’s qubit to, considering all possible measurements on her qubit. This steering
ellipsoid  can be used to give a faithful representation of an arbitrary two-qubit state in three dimensions. Moreover, in this scenario the core properties of the state and its
correlations are made manifest in simple geometric terms.
\par
Let $ \sigma_{\mu}=\{ \mathcal{I},\sigma_{x},\sigma_{y},\sigma_{z} \},~~\mu=0,1,2,3 $
denote the  \textit{Pauli basis}. Any two-qubit state $ \rho $ can be written
in the Pauli basis as 

          \begin{equation}\label{rhoPauli}
\rho=\frac{1}{4}\bigg(\sum_{\mu,\nu=0}^{3}\Theta_{\mu\nu} \sigma_{\mu}\otimes\sigma_{\nu}\bigg)
          \end{equation}
in which

          \begin{equation}\label{rhoPauli}
\Theta_{\mu\nu}=\text{Tr}\big(\rho~ \sigma_{\mu}\otimes\sigma_{\nu}\big) 
          \end{equation}
creates the elements of the block matrix $ \Theta=\left(
\begin{array}{ccc}
\mathcal{I} & \textbf{b}^{\top}  \\
\textbf{a} & T  \\
\end{array} \right) $
where $ \textbf{a},~ \textbf{b} $, whose elements are given by $ a_{j}=\text{Tr}\big( \rho~\sigma_{j}\otimes \mathcal{I} \big),~ b_{j}=\text{Tr}\big( \rho~\mathcal{I} \otimes \sigma_{j}  \big) $,   denote the Bloch vectors of
the reduced states $ \rho_{A} $ and  $ \rho_{B} $ of density matrix  $ \rho $, respectively, and $ 3 \times 3 $ matrix $ T $ 
 encodes the correlations. 
 \par
When Alice makes a projective measurement on her qubit, and
obtains an outcome corresponding to projector or positive-operator-valued measure (POVM) element $ E $, she steers Rob to  state $\rho _{B}^{ E}=\dfrac{\text{Tr}_{A}[\rho~E\otimes \mathcal{I}]}{\text{Tr}[\rho~E\otimes \mathcal{I}]} $ with probability $ P_{E}=\text{Tr}[\rho~E\otimes \mathcal{I}] $. 
Rob's steering ellipsoid $ \mathcal{E} $ 
gives the set of Bloch vectors to which Rob's qubit
can be collapsed considering all possible local measurements
performed by Alice. This ellipsoid has center \cite{JevticPRL,JevticJosza}

          \begin{equation}\label{center}
\text{\textbf{C}}=\dfrac{\textbf{b}-T^{\top}\textbf{a}}{1-a^{2}}
          \end{equation}
          and its semiaxes $ s_{1},~s_{2},~s_{3} $ are the roots of the eigenvalues of the following matrix

          \begin{equation}\label{ElipsoidQ}
Q=\dfrac{1}{1-a^{2}}\big(T^{\top}-\textbf{b}\textbf{a}^{\top}\big)\bigg(1+\dfrac{\textbf{a}\textbf{a}^{\top}}{1-a^{2}}\big(T-\textbf{a}\textbf{b}^{\top}\big)\bigg).
          \end{equation}
          \par
 The quantum coherence $ C $ of $ \rho _{B}^{ E} $ in the basis $ \{|\xi_{i}\rangle\} $ is defined as the summation of the absolute
values of off-diagonal elements \cite{Baumgratz}:
          \begin{equation}\label{coherence}
C\big(\rho _{B}^{ E},\{|\xi_{i}\rangle\}\big)=\sum_{i\neq j}|\langle \xi_{i} |\rho _{B}^{ E}|\xi_{j}\rangle|
          \end{equation}
Maximizing this coherence over all possible POVM operators $ E $ and  taking the infimum over all possible eigenbases $ \Xi $ for Rob, we can obtain
 the maximal steered coherence (MSC) of shared state $ \rho $ \cite{Hu2016}: 
            \begin{equation}\label{MAXcoherence}
 \Lambda(\rho)=\inf_\Xi \bigg\{\max_{E\in \text{POVM}}\big[  C\big(\rho _{B}^{ E},\{|\xi_{i}\rangle\}\big) \big]\bigg\}
           \end{equation}
           In Appendix \ref{A1} we obtain an analytical
           expression of MSC for X-states. 
\subsection{Quantum Fisher Information}
For a given quantum state $ \rho(X) $ parametrized by an
unknown parameter X, the unknown parameter may be inferred
from a set of measurements, usually described mathematically
by a set of POVM, on the quantum state.
By  optimizing  the measurements and the estimator, it is possible to obtain 
a precision limit of the unknown parameter estimation \cite{Braunstein1994}:

          \begin{equation}
Var(X)\geq \dfrac{1}{MF_{X}},
          \end{equation}
in which  $ M $ denotes the repeated times and $ F_{X} $ represents the
quantum Fisher information (QFI) of parameter $ X $ given by

          \begin{equation}\label{QFImohem}
F_{X}=\text{Tr}[\rho(X) L^{2}],
          \end{equation}
where L, called  the so-called symmetric logarithmic derivative (SLD), is a 
Hermitian operator, satisfying  equation
 
           \begin{equation}\label{SLD}
\partial_{X}\rho=\frac{1}{2}\{\rho(X),L\},
           \end{equation}
  where  $ \{...\} $ stands for the anti-commutator. 
Taking the trace on both sides of above equation, we can see that $  \langle L \rangle=\text{Tr}\{\rho L\}=0$. Hence, the QFI
is actually the variance of the SLD operator.
 Considering the spectral decomposition of  density matrix $ \rho $ as  $ \sum\limits_{i=1}^{M}P_{i}|\psi_{i}\rangle\langle \psi_{i}| $ where $ M$ 
represents the dimension of  the support of $ \rho $ and $|\psi_{i}\rangle  $'s
  ($ P_{i} $'s)   denote   the  eigenstates (nonzero eigenvalues) of $ \rho $, one can write the elements of the SLD operator as \cite{J. Liu11}.

\begin{equation}
L_{ij}=\dfrac{\partial_{X}P_{i}}{P_{i}}\delta_{ij}+\dfrac{2(P_{i}-P_{j})}{P_{i}+P_{j}}\langle \partial_{X}\psi_{i} |\psi_{j}\rangle,
\end{equation}
where  $ i,j\in [1,M] $. Moreover, for $ i $ and $ j $,  larger than $ M $,  $L _{ij} $ can be an arbitrary number. It should be noted when $ \rho $ is positive definite (or full rank), $ M $ equals the dimension of the Hilbert space. Finally, it is possible to obtain the following  expression of QFI for a non-full rank density matrix \cite{J1,J2,J3}

\begin{equation}
F_{X}=\sum\limits_{i=1}^{M}\dfrac{(\partial_{X}P_{i})^{2}}{P_{i}}+\sum\limits_{i=1}^{M}4P_{i}\langle \partial _{X} \psi_{i}| \partial _{X} \psi_{i}\rangle -\sum\limits_{i,k=1}^{M}\dfrac{8P_{i}P_{k}}{P_{i}+P_{k}}|\langle  \psi_{i}| \partial _{X} \psi_{i}\rangle|^{2}.
\end{equation}

It should be noted that the above formula  can cover the full rank case when choosing $ M =d $, where $ d $ represents the Hilbert space dimension. 
Besides, the QFI may be also rewritten as follows

\begin{equation}
F_{X}=\sum\limits_{i=1}^{M}\dfrac{(\partial_{X}P_{i})^{2}}{P_{i}}+\sum\limits_{i=1}^{M}4P_{i}F_{X,i} -\sum\limits_{i\neq k}^{M}\dfrac{8P_{i}P_{k}}{P_{i}+P_{k}}|\langle  \psi_{i}| \partial _{X} \psi_{i}\rangle|^{2},
\end{equation}
 where $ F_{X,i} $ denotes  the QFI for ith pure eigenstate:
 
  \begin{equation}
  F_{X,i}=4\big(\langle \partial _{X} \psi_{i}| \partial _{X} \psi_{i}\rangle-|\langle  \psi_{i}| \partial _{X} \psi_{i}\rangle|^{2}\big).
  \end{equation}
  
  \subsection{Weak measurement and weak measurement reversal}
  
The unsharp measurement  being considered in this paper, is the weak or partially collapsed
measurement examined  in Refs. \cite{Korotkov2006,Neeley2008}.
An important part of the partial measurement (PM) is a detector measuring the qubit and functioning as
follows: the detector clicks with a probability $ p $ provided that the qubit is in  $ |1\rangle $ state and never clicks
if the qubit is in  $ |0\rangle $ state. Therefore, the detector provides some partial information about the
initial state of the qubit \cite{KimExpress}. 

\par

We first  assume that the detector has clicked. This  is identical to the normal
projection measurement in which the  qubit state is irrevocably collapsed to $ |0\rangle $ state.
The measurement operator describing this situation may be written as follows:

\begin{equation}\label{measur1}
M_{1}=\sqrt{p}|0\rangle\langle 0|=\begin{bmatrix}
    \sqrt{p} & 0  \\
    0 & 0
  \end{bmatrix},
\end{equation}
Because $M _{1} $ has no mathematical inverse, it is not reversible and hence $ M_{1} $ is of no interest to us.

\par
Let us now assume that detector has not clicked. Therefore, the measurement operator
$ M_{0} $, corresponding to  the null output of the detector, may be evaluated by using the relation $ \mathcal{I}=M^{\dagger}_{0}M_{0}+M^{\dagger}_{1}M_{1} $
and can be written as

\begin{equation}\label{measur2}
M_{0}=\sqrt{1-p}|0\rangle\langle 0|+|1\rangle\langle 1|=\begin{bmatrix}
    \sqrt{1-p} & 0  \\
    0 & 1
  \end{bmatrix}.
\end{equation}
This is exactly the PM at the focus of this paper. The variable $ p $, defined as the partial-collapse strength,  corresponds to p = 1 for the  normal projection measurement. In order to reverse the PM effect, i.e.,  recovering the initial (original) state $ \rho_{i} $ from the post
measurement state  given by $ \rho_{f}=\dfrac{M_{0}\rho_{i}M^{\dagger}_{0}}{\text{Tr}\big(M_{0}\rho_{i}M^{\dagger}_{0}\big)} $, we only need to apply
the inverse of $ M_{0} $, obtained by replacing $ q $ with $ p $ in the following expression:

\begin{equation}\label{measur3}
M_{0}^{-1}=\left(\begin{array}{cc}
\dfrac{1}{\sqrt{1-q}}& 0 \\
 0 & 1 \\
\end{array}\right)
=\frac{1}{\sqrt{1-q}}\sigma_{x}M_{0}\sigma_{x}=\dfrac{1}{\sqrt{1-q}}M^{rev}_{0},
\end{equation}
referring to this point that the reversing operation $M^{rev}_{0}  $ is implemented by the sequence of a
bit-flip operation $ \sigma_{x} $, another weak measurement $M _{0} $, and a final bit-flip operation. Note that we have assumed the 
more general case that the reversing operation  may be  applied with  different strength $ q $.

\section{The Model   \label{Model}}
\begin{figure}[ht]
 \includegraphics[width=8cm]{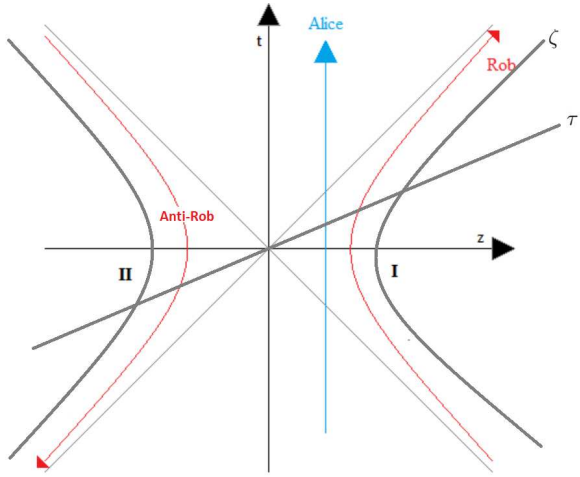}
 \caption{\small Minkowski spacetime and Rindler coordinate }
 \label{f1}
   \end{figure}
We consider  two
   modes of a Dirac field described by relatively accelerated
      parties; an inertial observer Alice (A) and a uniformly accelerated observer
Rob (R) moving with a constant acceleration $ a $.
Each of the two parties is  assumed to possess a
detector sensitive only to one of the two modes. 
We focus on the Unruh effect for Dirac particles
as experienced by Rob \cite{Alsing2006}. When a given Dirac mode is in the
vacuum state from an inertial perspective,  Rob's detector perceives a Fermi-Dirac distribution of particles.
  Alice moves in the
Minkowski plane with the coordinates $ (t, z) $, as shown in Fig. (\ref{f1}). The Minkowski coordinates are the most suitable to describe
the field from an inertial perspective. However,
a uniformly accelerated observer is unable to access information about the whole of spacetime, because  a communication horizon appears
from his perspective. Hence, for  formulating this phenomenon, 
the setting of the constant acceleration may
be conveniently described by the Rindler coordinate $ (\tau, \zeta) $ involving two disconnected
regions I and II. In this coordinates, 
one can describe the uniformly accelerated Rob to travel on a
hyperbola constrained to region I, as shown in  Fig. (\ref{f1}). In fact,  Rob has no access to the field modes
in the causally disconnected region II. Hence, he must trace over
the inaccessible region II, leading to an unavoidable loss of
information about the state and  essentially resulting in the
detection of a mixed state. Under the single mode approximation, the Minkowski vacuum state $ |0\rangle_{M} $ and
the only excited state
(one-particle state) $ |1\rangle_{M} $
may be expressed  in
terms of the Rindler regions I and II states:

\begin{equation}\label{M1}
|0 \rangle_{M}=\text{cos}~r|0\rangle_{I}|0\rangle_{II}+\text{sin}~r|1\rangle_{I}|1\rangle_{II},\\
\end{equation}
\begin{equation}\label{M2}
|1 \rangle_{M}=|1\rangle_{I}|0\rangle_{II}
\end{equation}
where the dimensionless acceleration parameter $ r $ is defined by $ r=\text{arccos}\sqrt{1+e^{\frac{-2\pi\omega}{a}}} $ in which $ \omega $ and $a  $ are   the Unruh mode frequency and   acceleration, respectively, with  $ 0< a<\infty $ and hence $ r\in\left[ 0,\pi/4\right] $.

\par
In order to investigate how the measurement before and after  accelerating Rob, affects the process of parameter estimation, we consider the following scenario \cite{Xiao2017}: 

(i) Initially, Alice
and Rob share the  entangled state
\begin{equation}\label{step1}
|\Psi\left( 1\right) \rangle=\text{sin}\left(\frac{\vartheta}{2}\right)|0\rangle_{A}|0\rangle_{R}+\text{cos}\left(\frac{\vartheta}{2}\right)~\text{e}^{i\varphi}|1\rangle_{A}|1\rangle_{R};~~\ 0\leq\vartheta\leq\pi,\ 0\leq\varphi\leq 2\pi.
\end{equation}

(ii) A PM is performed by Rob
 on his own particle before the acceleration  (i.e., at time  $ \tau=0 $).

 (iii) We assume that Rob  undergoes uniform acceleration $ a $ and hence states $ |0\rangle_{R} $ and $ |1\rangle_{R} $ should be expanded as Eqs. (\ref{M1}) and (\ref{M2}), respectively.

(IV) After Rob's acceleration, the operation of partial measurement
reversal (PMR) is implemented by Rob in the region I.

 Assuming that the measurements have been performed successfully, we obtain the following mixed
state between Alice and Rob after tracing over region II:

\begin{eqnarray}\label{a16}
\rho_{A,R}=\frac{1}{N}\left(\begin{array}{cccc}
\text{sin}^{2}\frac{\vartheta}{2}~\overline{p}~\text{cos}^{2}r& 0& 0 & \text{sin}~\frac{\vartheta}{2}\text{cos}~\frac{\vartheta}{2}~e^{-i\varphi}~\sqrt{\overline{pq}}~\text{cos}r\\
 0 & \text{sin}^{2}\frac{\vartheta}{2}~\overline{pq}~\text{sin}^{2}r&0 & 0 \\
 0 & 0& 0 &0 \\
\text{sin}~\frac{\vartheta}{2}\text{cos}~\frac{\vartheta}{2}~e^{i\varphi}~\sqrt{\overline{pq}}~\text{cos}r & 0 & 0& \text{cos}^{2}\frac{\vartheta}{2}~\overline{q} \\
\end{array}\right),~~~
\end{eqnarray} 
where  $ N=\text{sin}^{2}\frac{\vartheta}{2}~\overline{p}~\text{cos}^{2}r+\text{sin}^{2}\frac{\vartheta}{2}~\overline{pq}~\text{sin}^{2}r+\text{cos}^{2}\frac{\vartheta}{2}~\overline{q} $ is the normalization factor and $ \overline{p}=1-p $ as well as $ \overline{q}=1-q $.

\section{Premeasurement and postmeasurement effects on the behaviour of QFI \label{PE}}
 \subsection{Weight parameter estimation}
 
  \begin{figure}[ht!]
  \subfigure[]{\includegraphics[width=6.5cm]{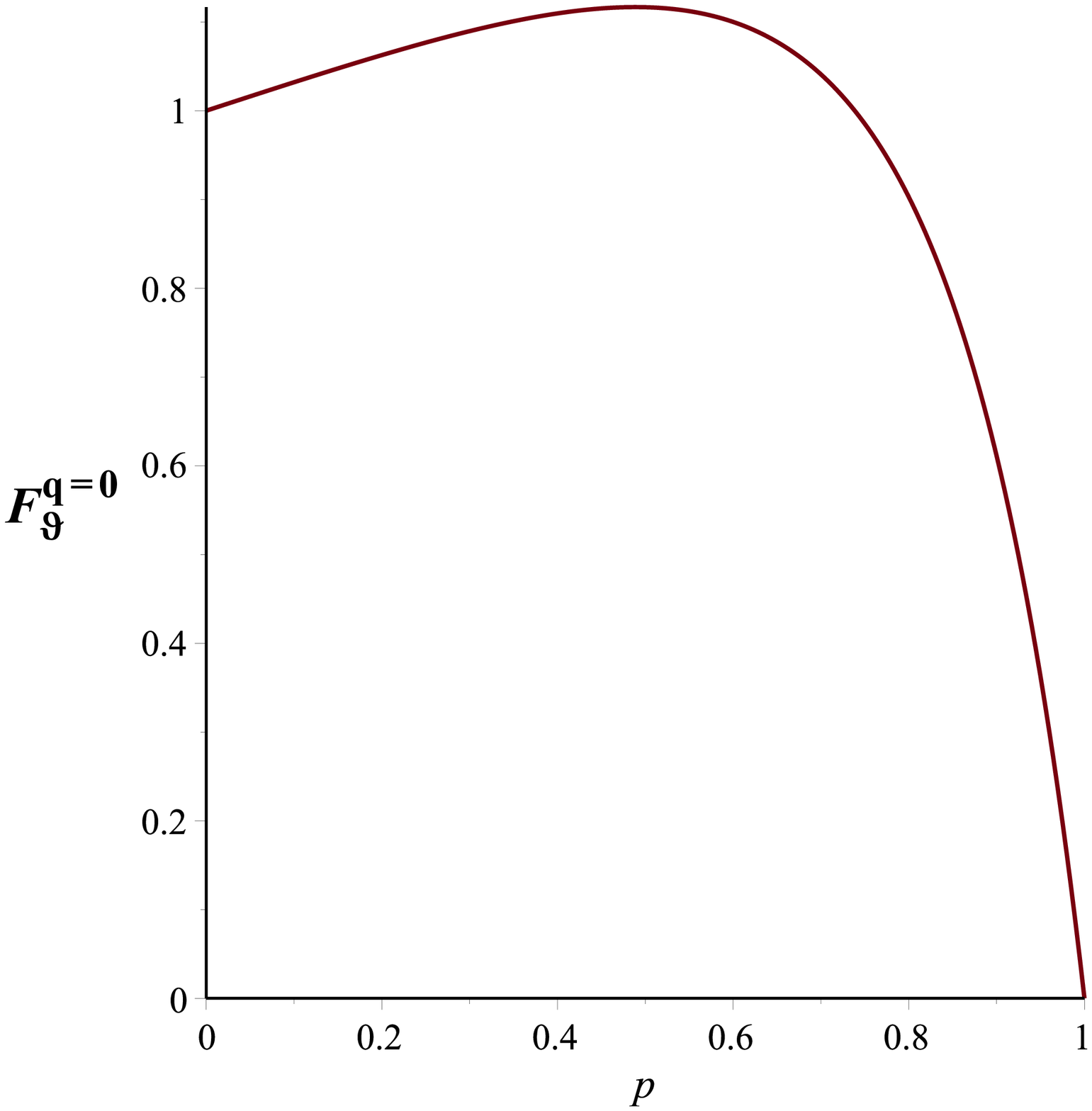}\label{Ftetaqzero1} }
  \subfigure[]{\includegraphics[width=6.5cm]{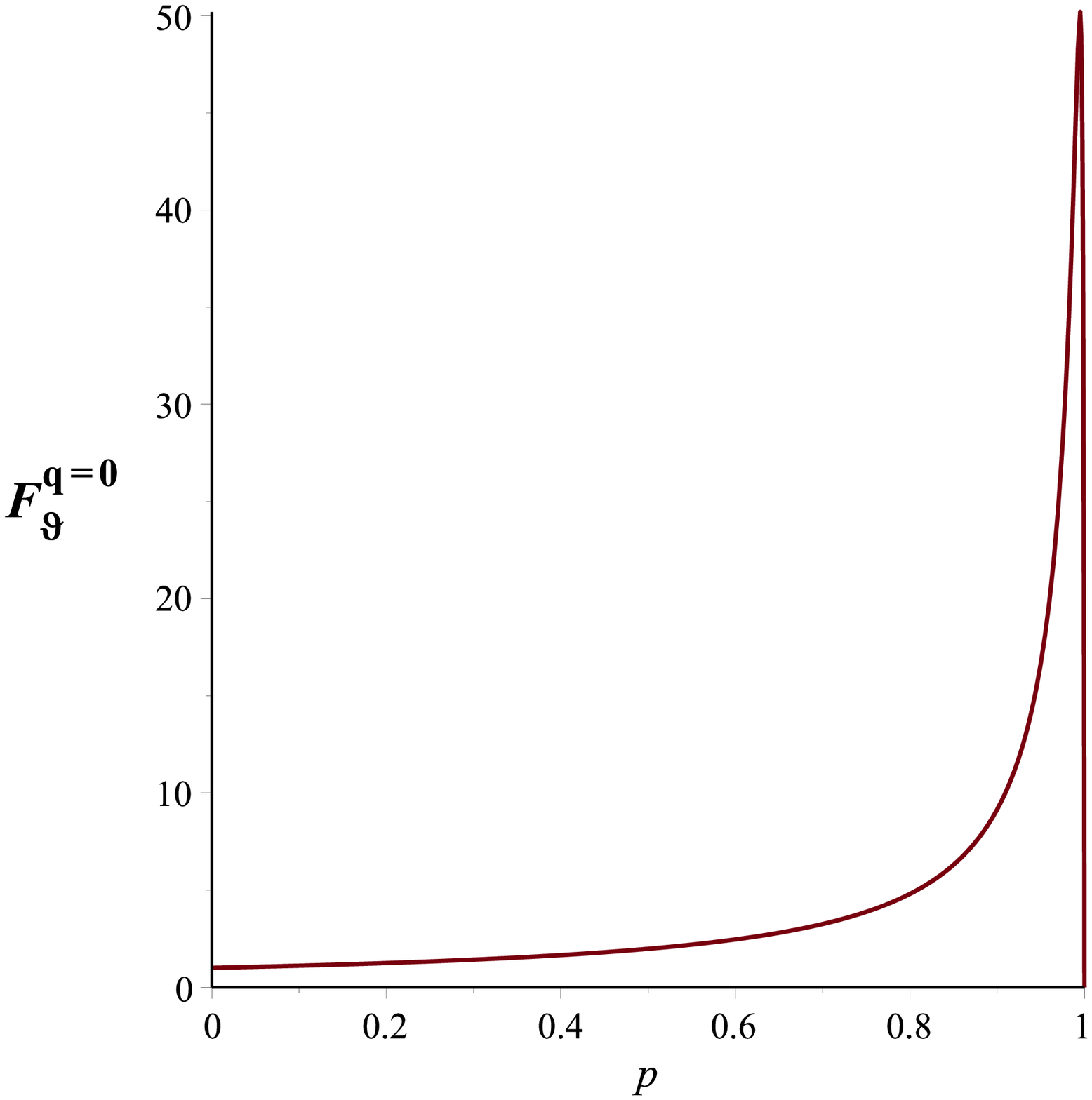}\label{Ftetaqzero2} }
   \caption{The QFI corresponding to the estimation of  weight parameter  for (a) $  \vartheta =1.9 $ and (b) $  \vartheta =3 $  when no post-measurement is performed. }
     \end{figure}

     \begin{figure}
     \begin{minipage}[ht]{0.5\linewidth}
     \centering
     \includegraphics[width=6.5 cm]{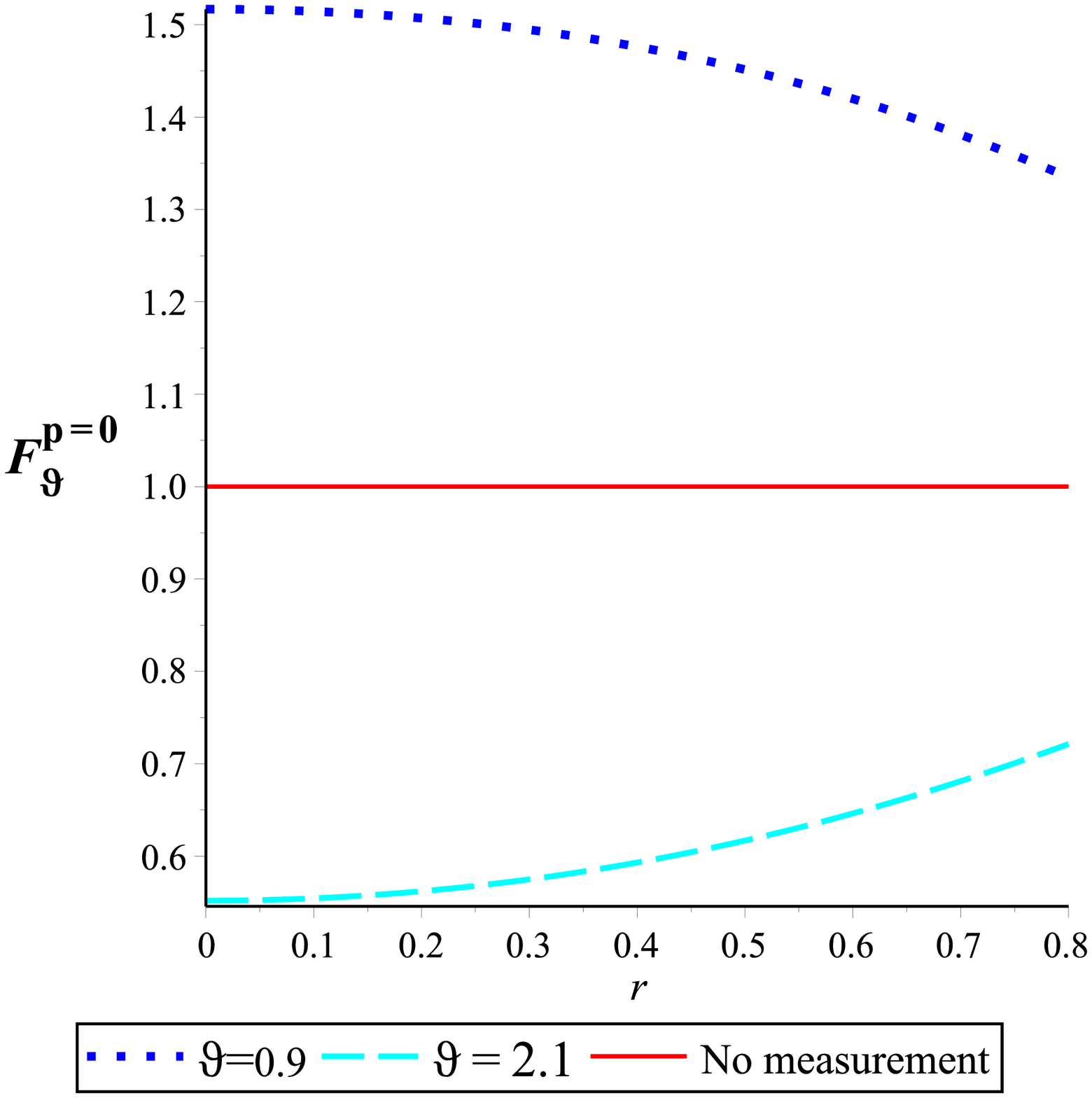}
     \caption{{ {\scriptsize The QFI corresponding to the estimation of  weight parameter  for $ q=0.6 $ and different values of $ \vartheta $  when only the postmeasurement is made.}}}
     \label{Ftetapzero}
     \end{minipage}%
     \begin{minipage}[ht]{0.5\linewidth}
     \centering
           \subfigure[]{\includegraphics[width=6.5cm]{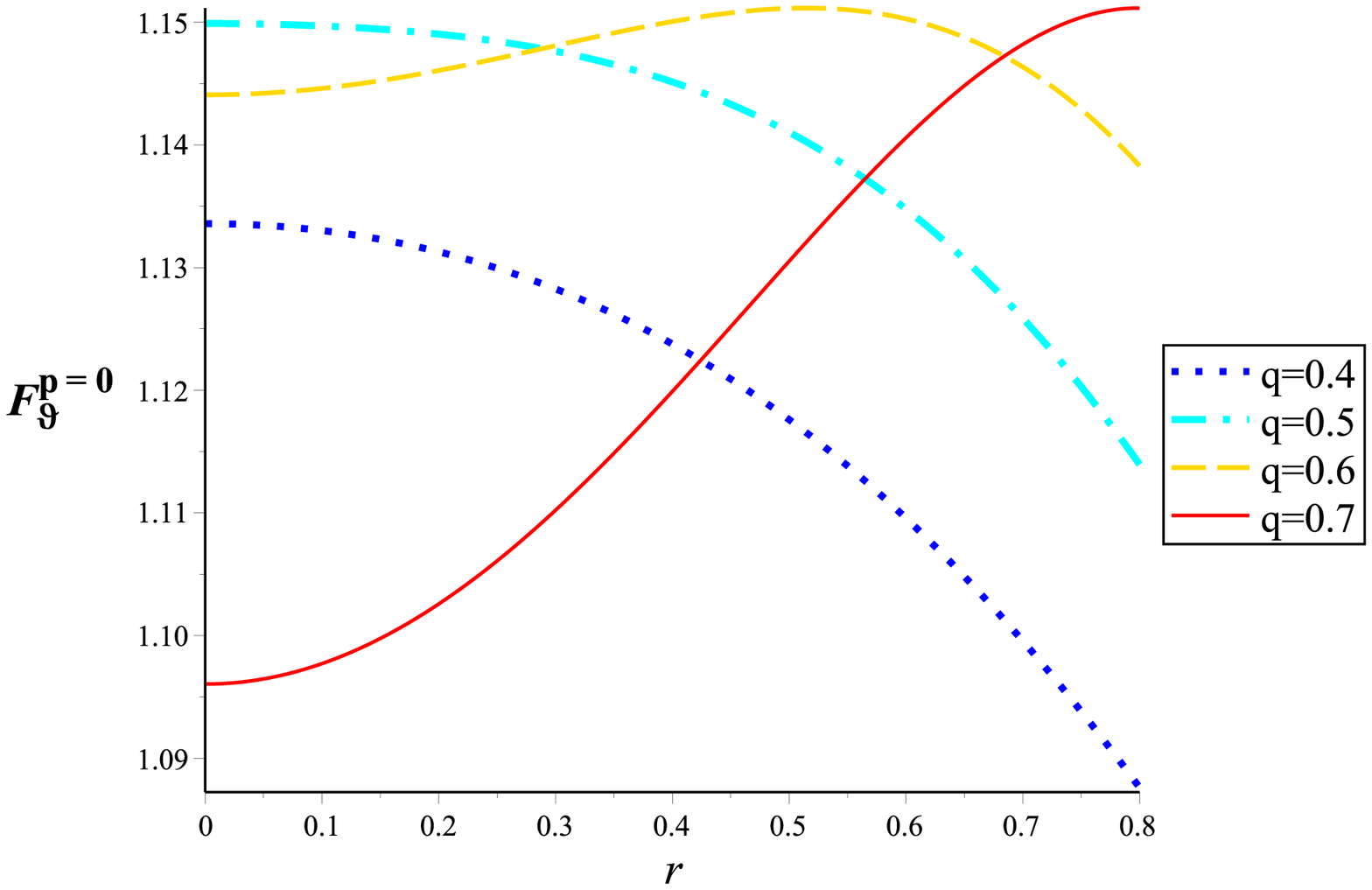}\label{Ftetapzeroq1} }
           \subfigure[]{\includegraphics[width=6.5cm]{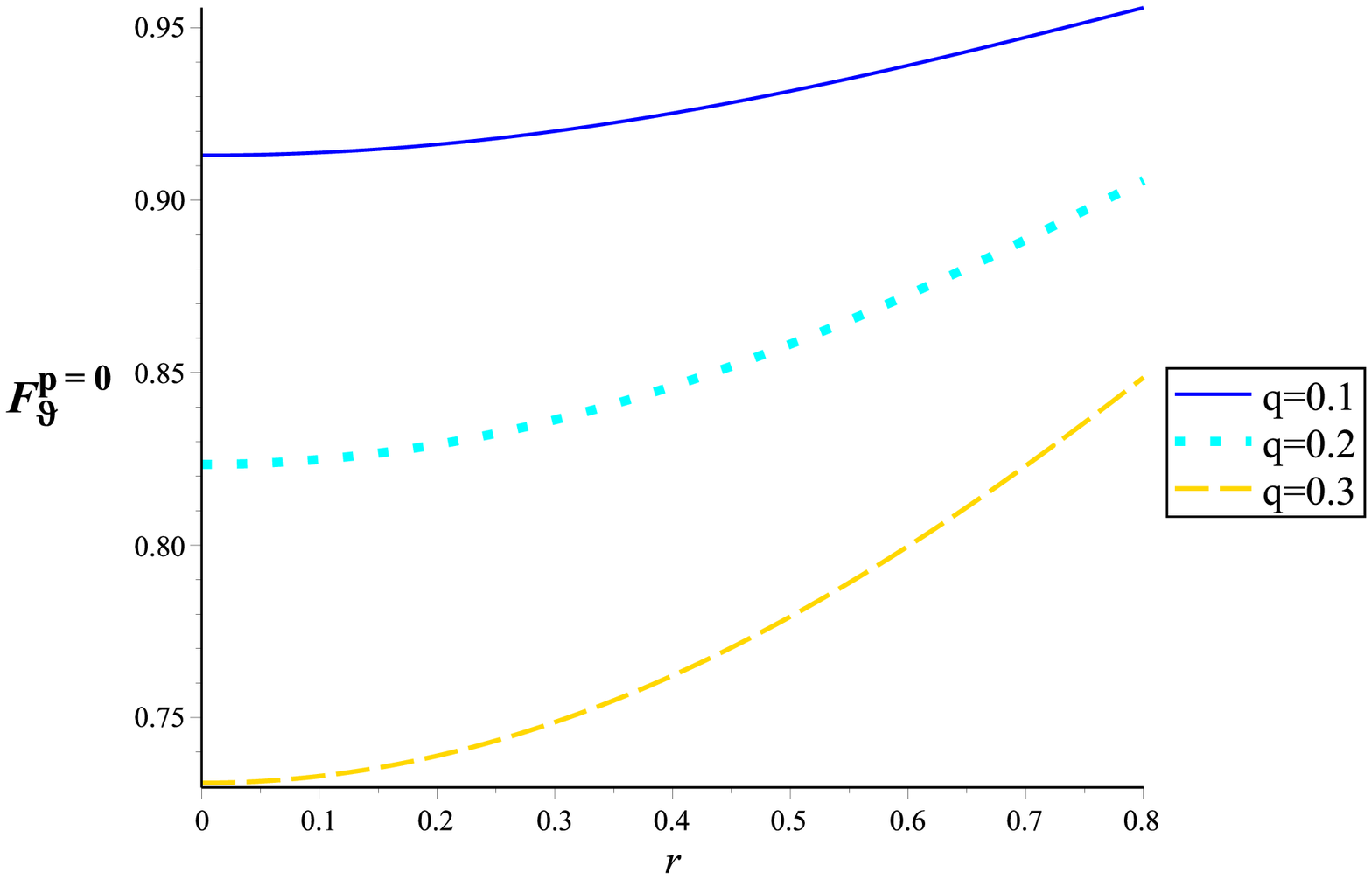}\label{Ftetapzeroq2} }
     \caption{{\scriptsize The QFI corresponding to the estimation of  weight parameter  for (a) $  \vartheta =1.2 $ as well as (b) $  \vartheta =2.6 $  and different values of $ q $ when only the postmeasurement is made.}} \label{Ftetapzeroq}
     \end{minipage}
     \end{figure}

       \begin{figure}[ht]
             \includegraphics[width=10cm]{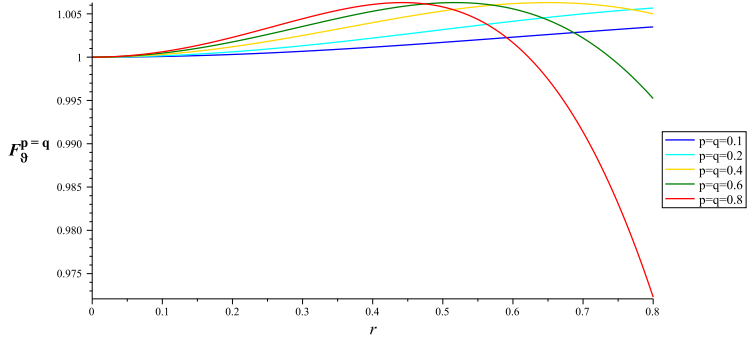}
             \caption{\small   The QFI versus $ r $ for $ \vartheta> \pi/2~ ( \vartheta=1.65) $ when both measurements are made with equal strength.}
             \label{Fpq}
               \end{figure}
               
                \begin{figure}[ht!]
                     \subfigure[]{\includegraphics[width=4.8cm]{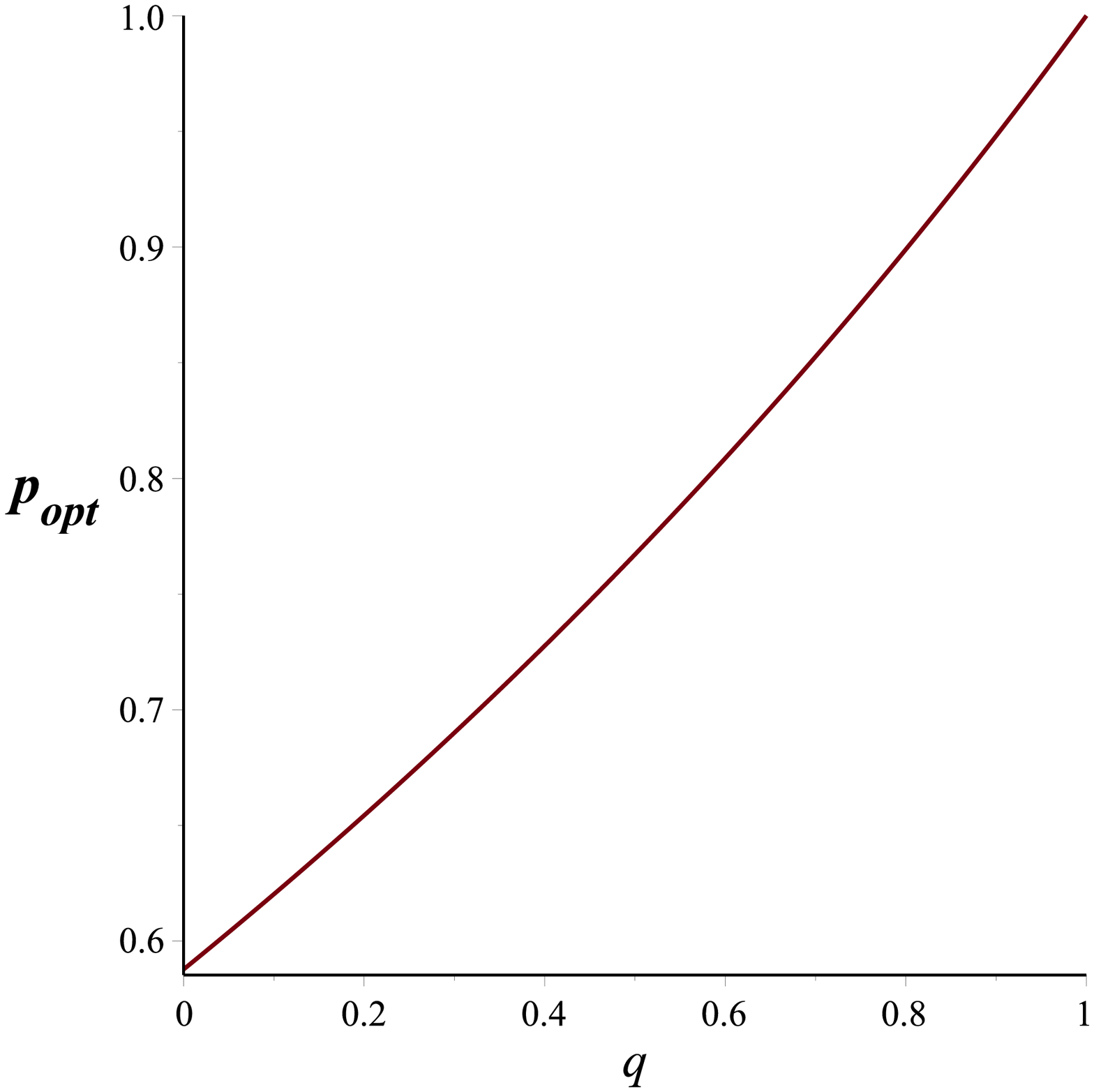}\label{POPTq} }
                     \subfigure[]{\includegraphics[width=4.8cm]{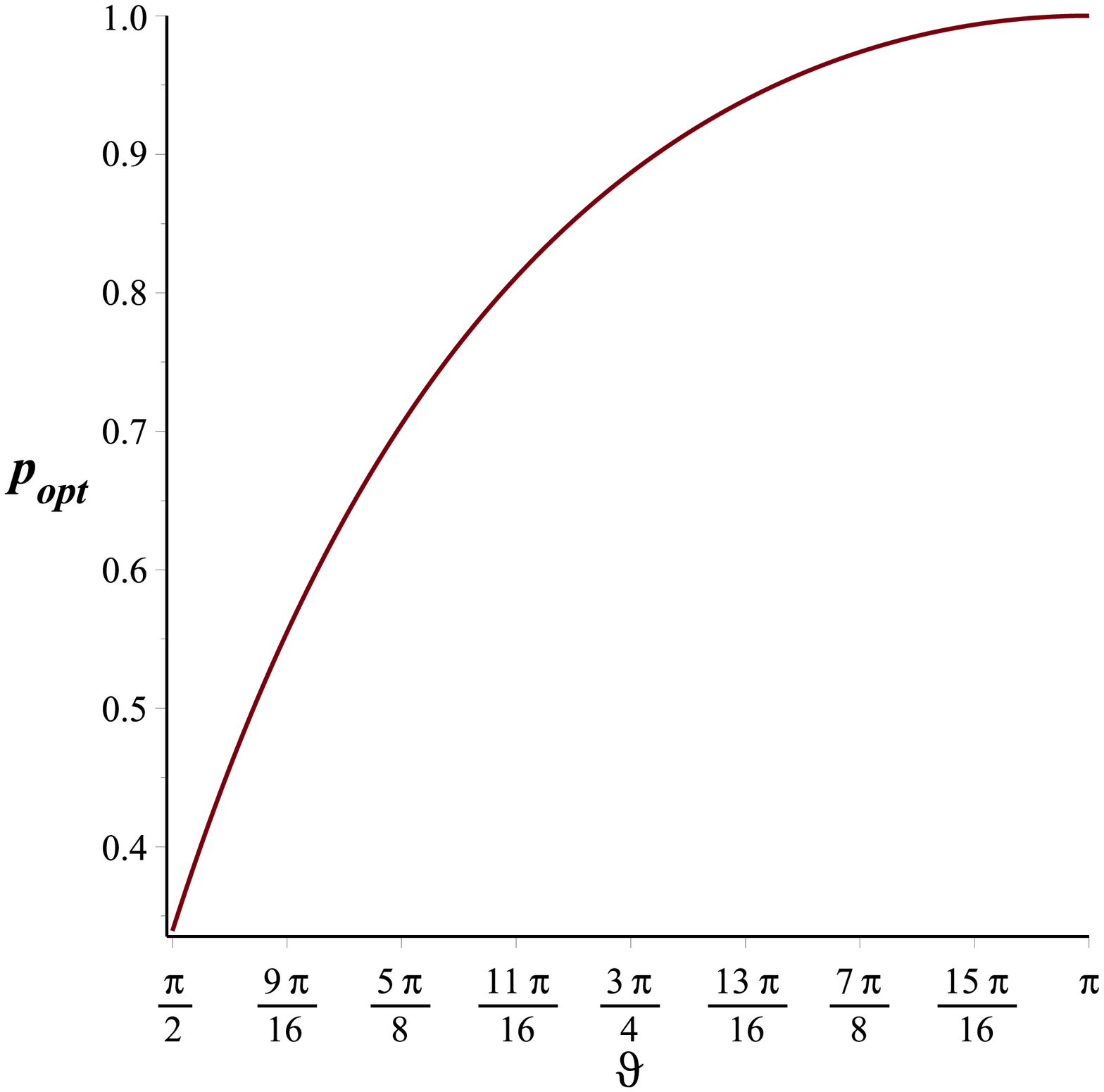}\label{POPTvar} }
                     \subfigure[]{\includegraphics[width=4.8cm]{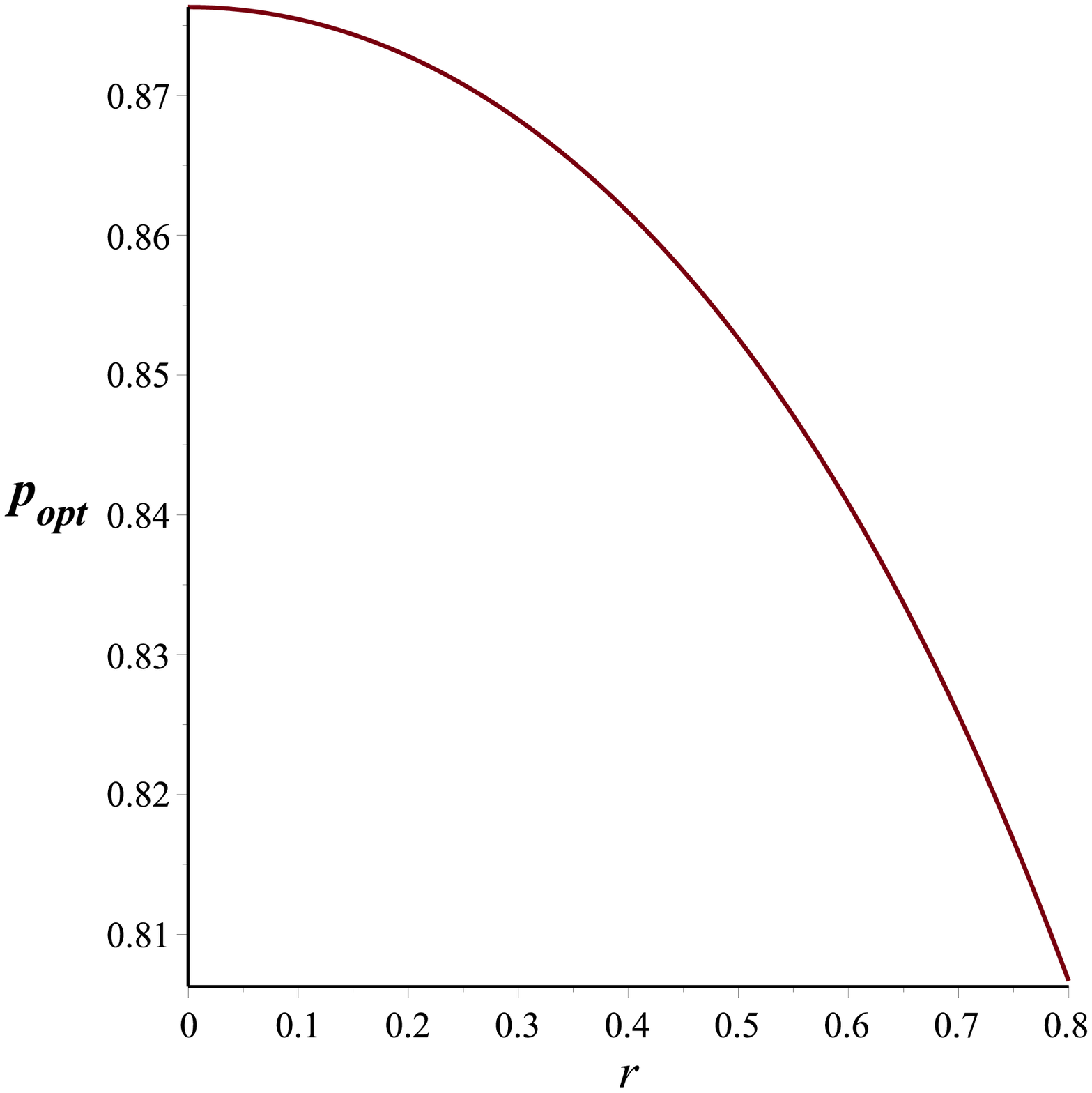}\label{POPTr} }
                     \caption{\small The optimal value of premeasurement strength  as functions of $ q $, $ \vartheta $, and $ r $ for (a) $ \vartheta=2 $ as well as $ r=0.5 $; (b) $ q = 0.4 $ as well as  $ r = 0.5 $, and (c) $ q=0.7 $ as well as $ \vartheta=0.2 $}
                     \label{POPTfig}
                       \end{figure}
                       
                        \begin{figure}[ht!]
       \subfigure[]{\includegraphics[width=4.8cm]{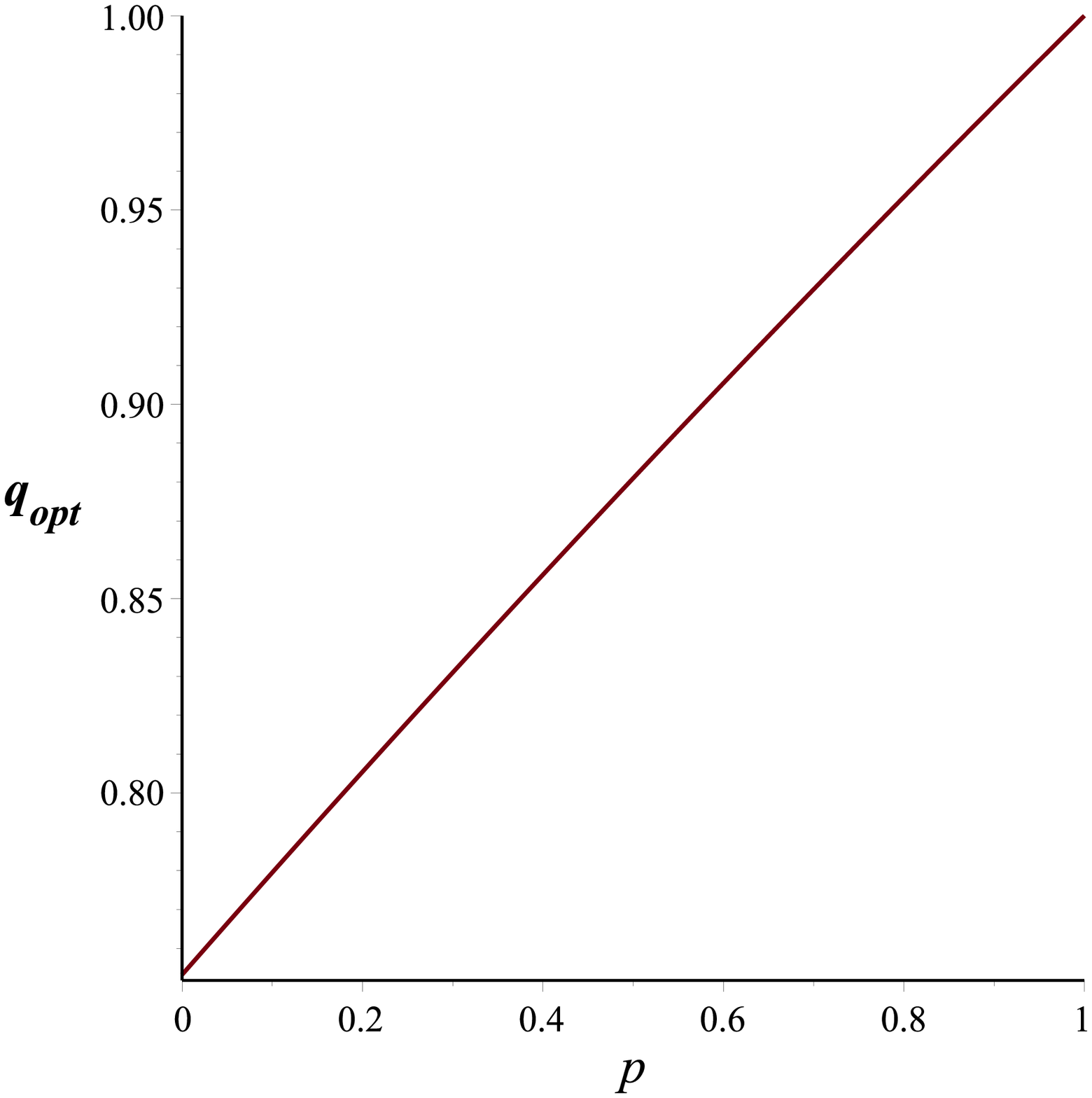}\label{QOPTp} }
       \subfigure[]{\includegraphics[width=4.8cm]{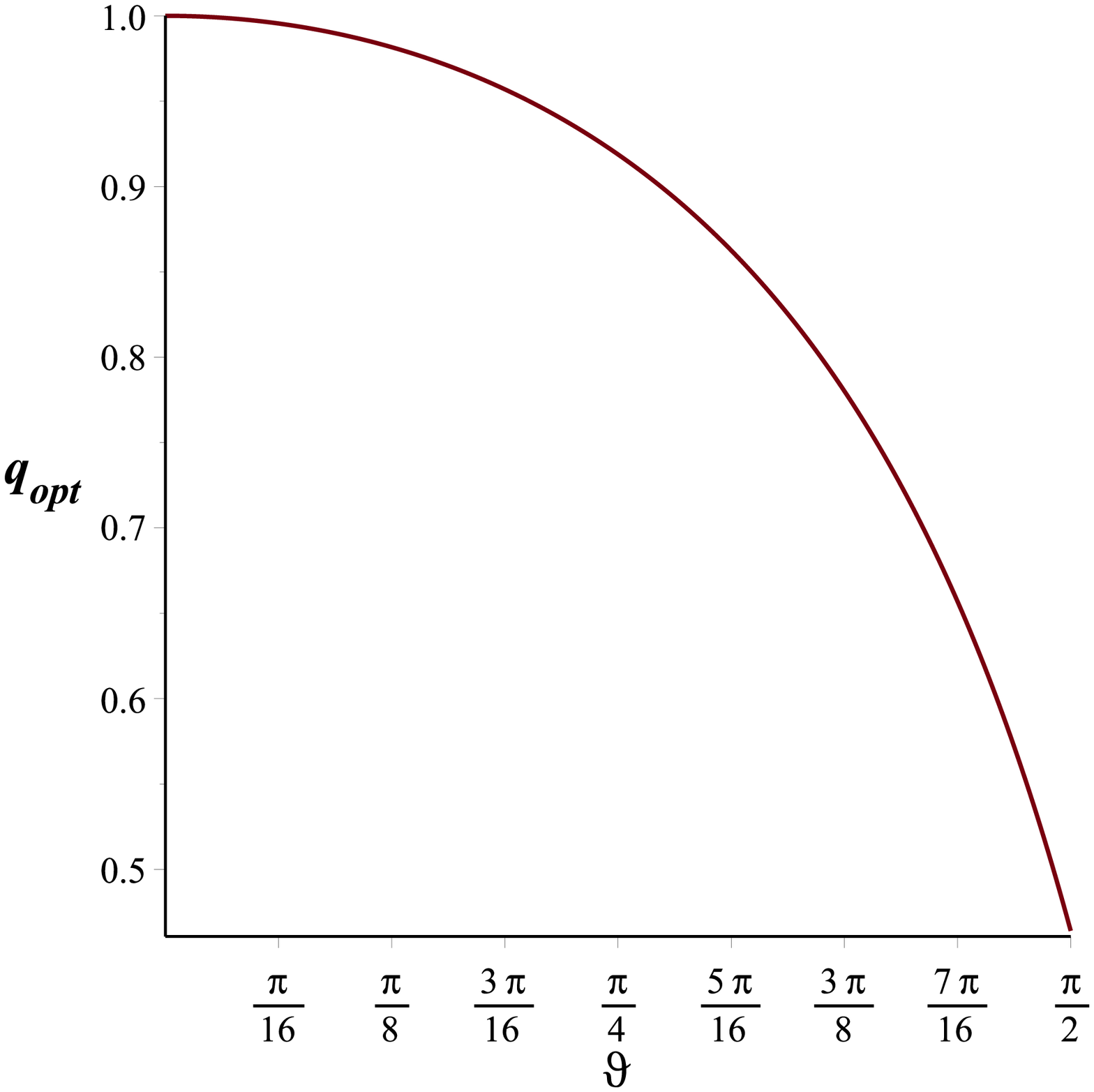}\label{QOPTvar} }
       \subfigure[]{\includegraphics[width=4.8cm]{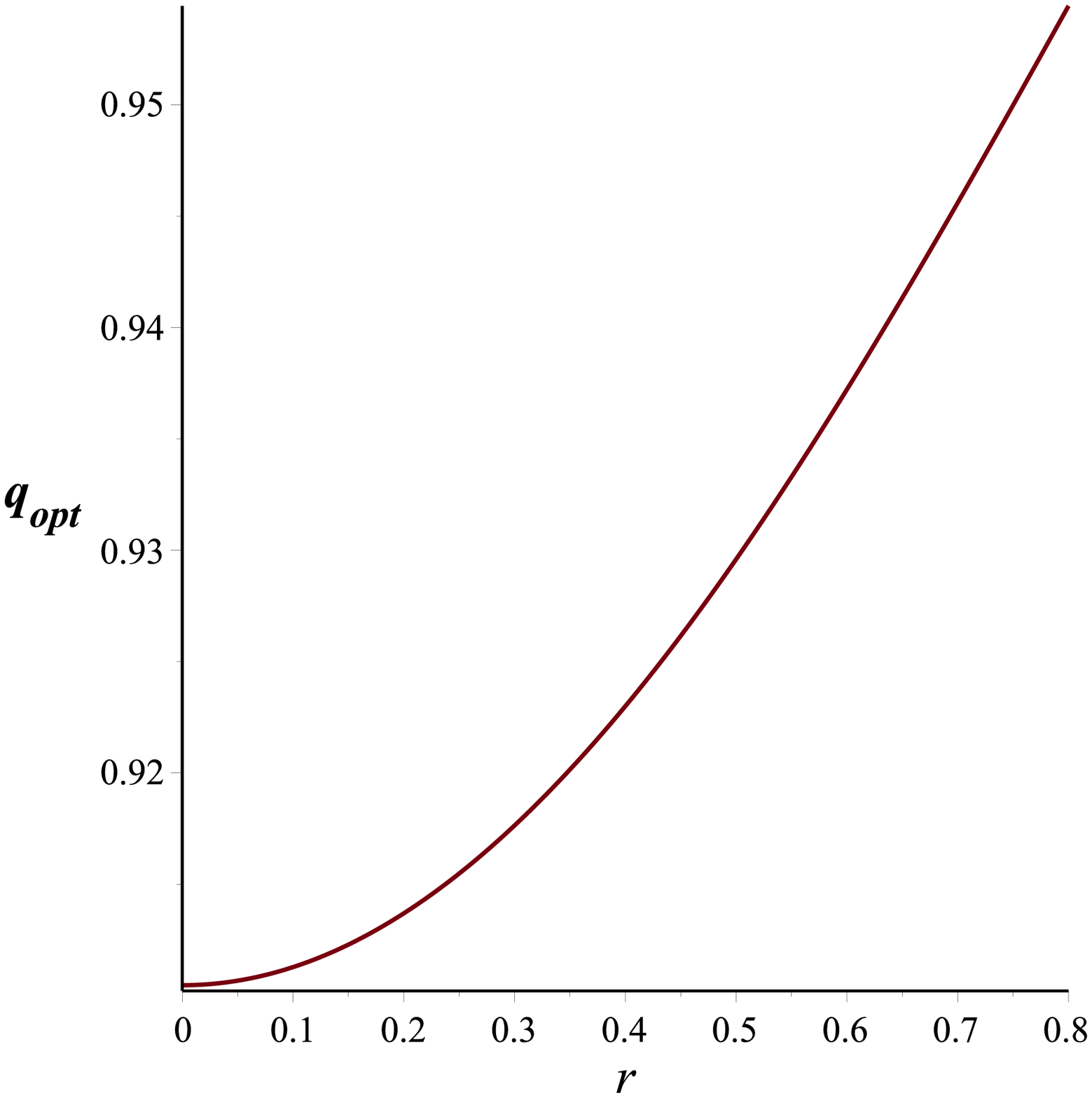}\label{QOPTr} }
           \caption{\small The optimal value of postmeasurement strength  as functions of $ p $, $ \vartheta $, and $ r $ for (a) $ \vartheta=1 $ as well as $ r=0.5 $; (b) $ p = 0.4 $ as well as  $ r = 0.5 $, and (c) $ p=0.7 $ as well as $ \vartheta=1 $}
                                            \label{QOPTfig}
                                              \end{figure}
  In this section, we focus on the estimation of the weight parameter $ \vartheta $. For calculation of the QFI, we use the following method \cite{RanganiOPTC}.
   First, the block diagonal  state (\ref{a16}) should be written in the form  $ \rho_{A,I}=\bigoplus^{n}_{i=1}\rho_{i}$, in which $ \bigoplus $ represents the direct sum. Then, it  can be checked that the SLD operator may be  written as  $L=\bigoplus^{n}_{i=1}L_{i}$, where $L_{i}$ indicates the corresponding SLD operator for $\rho_{i}$.
    It can be shown that   the SLD operator for the $i$th block is given by \cite{J. Liu11}
    \begin{equation}
      L_{i}=\frac{1}{\mu_{i}}\left[\partial_{x}\rho_{i}+\xi_{i}\rho^{-1}_{i} -\partial_{x}\mu_{i} \right],
       \end{equation}
    where $\xi_{i}=2\mu_{i}\partial_{x}\mu_{i}-\partial_{x}P_{i}/4$ in which $\mu_{i}=\text{Tr}\rho_{i}/2$ and $P_{i}=\text{Tr}\rho^{2}_{i}$. Note that $\xi_{i}$ vanishes if det$ \rho_{i}=0 $. 
    
    \par
     Constructing SLD $ L $ with this method
     and inserting  it into Eq. (\ref{QFImohem}) with   density matrix (\ref{a16}), we obtain the following expression for the QFI  {associated with weight parameter $ \vartheta $:

  \begin{equation}\label{qfivar}
F_{\vartheta} =\,{\frac {-8~\overline{q}\overline{p} \left( -\overline{q}+ \left( \overline{q}-1 \right) \cos \left( 2\,r \right) 
 -1 \right) }{ \big[ - \left( 2\,\overline{q}-2 \right) \overline{p}\cos \left( 2\,r
  \right)   \sin^{2} \left( \vartheta /2 \right)  +\overline{p}+\overline{qp}+2
 \,\overline{q}- \left( \overline{qp}+\overline{p}-2\,\overline{q} \right) \cos \left( \vartheta  \right)  \big] 
 ^{2}}}.
  \end{equation}
  We find when $ p,q\rightarrow 0 $, then $ F_{\vartheta}\rightarrow 1 $. Therefore, when no measurement is performed, the Unruh decoherence and the initial state do not affect on the weight parameter estimation. Especially, when the  post-measurement does not perform, i.e., $ q=0 $, the QFI is reduced to:
  
  \begin{equation}\label{qfivarp}
  F^{q=0}_{\vartheta} =
{\frac {4-4\,p}{ \left( \cos \left( \vartheta  \right) p-p+2 \right) ^
{2}}},
    \end{equation}
  denoting that the QFI is unaffected by the Unruh effect. Hence, the second measurement plays a key role for determining 
  whether or not  the estimation of the weight parameter is affected by the Unruh effect. Now investigating (\ref{qfivarp}), it is observed that for $ \vartheta < \pi/2 $, the QFI degrades with increasing $ p $, while for $ \vartheta > \pi/2 $, 
  the QFI may increase with $ p $ (see Fig. \ref{Ftetaqzero1}). Especially, as seen in Fig. \ref{Ftetaqzero2}, when $ \vartheta\rightarrow \pi $, the estimation of the weight parameter is enhanced considerably with increase in the pre-measurement strength, unless $ p\rightarrow 1 $, in that case the 
  QFI decreases with a steep slope. Therefore, the PM  may guarantees enhancement of the weight parameter estimation, however approaching the 
  sharp  von Neumann measurement sleeply decreases the precision of the estimation. In order to obtain the optimum value of the PM strength for the best estimation, in the absence of the postmeasurement, we derive  (\ref{qfivarp}) in terms of $ p $. The result is as follows:
  
  \begin{equation}\label{popt}
    p_{q=0}^{\text{opt}}={\frac {2\cos \left( \vartheta  \right) }{\cos \left( \vartheta 
     \right) -1}},
      \end{equation}
  leading to the optimal QFI 
  
  \begin{equation}\label{qopt}
     (F^{q=0}_{\vartheta})_{\text{opt}}=\dfrac{1}{\text{sin}^{2}\vartheta}.
        \end{equation}
  
  Figure \ref{Ftetapzero} illustrates the pure effects of the post-measurement on the QFI and it is  compared with the case that no measurement is performed. The second PM, as described above, plays the role of a quantum key for manifestation of  the Unruh effect. When only the second measurement is performed, i.e., $ p=0 $, we see if  $ \vartheta < \pi/2 $, 
  the accuracy of the parameter estimation is improved compared to the situation in which no measurement has been made $(F_{\vartheta}=1)$, but when $ \vartheta > \pi/2 $, it is not possible to achieve a better estimation. Moreover, similar to the previous discussion,  approaching the 
    complete  von Neumann measurement,  $ q\rightarrow 1 $, causes the precision of the estimation to diminish and   be less than unity.
   
   In Fig. \ref{Ftetapzeroq}, we analyse how the postmeasurement  affects  the QFI behaviour versus the
    Unruh effect in the absence of the premeasurement. As seen in  Fig. \ref{Ftetapzeroq}(a), if   $ \vartheta < \pi/2 $, for small values of $ q $, the QFI decreases with growth of the acceleration parameter $ r $. Nevertheless,  with strengthening the postmeasurement, the precision of the estimation may be enhanced with increase in the acceleration and then it decreases. The optimal point is as follows:

      \begin{equation}\label{ropt}
        r_{p=0}^{\text{opt}}=\frac{1}{2}\,\arccos \left({\frac { \left( 4\,q-8 \right)  \sin^{2}
         \left( \vartheta /2 \right) -q\cos \left( \vartheta 
         \right) -3\,q+4}{2q \sin^{2} \left( \vartheta /2 \right)  }} \right).
          \end{equation}
    Inserting it in Eq. (\ref{qfivar}) with $ p=0  $, we can obtain the following expression for the optimal QFI:
    \begin{equation}\label{Qopt}
         (F^{p=0}_{\vartheta})_{\text{opt}}=\dfrac{1}{\text{sin}^{2}\vartheta}.
            \end{equation}
    
Moreover, for large values of $ q $, the QFI monotonously increases as the Unruh acceleration raises.  Similarly, when $ \vartheta > \pi/2 $, as illustrated in Fig. \ref{Ftetapzeroq}(b), the QFI and hence the precision of estimation monotonously are enhanced with increase in the acceleration.

  \par
  Now we focus on behaviour of the QFI when both measurements are made simultaneously. In the special case that  $ \vartheta > \pi/2 $ and $ p=q $
  the QFI may increase compared it to the situation that no measurement is performed (i.e., it is possible to obtain an optimal value or an enhanced value for the QFI such that $ F^{p=q}_{\vartheta}>1 $ ). Figure \ref{Fpq} shows the QFI as a function of  the Unruh acceleration parameter for $ p=q $. Generally,
   the maximum point of the plot of QFI versus $ r $ is given by:
  
   \begin{equation}\label{generalROPT}
           r_{\text{opt}}=\frac{1}{2}\,\arccos \left( {\frac { \left(  \left( q-2 \right) p-3\,q+4
            \right)   \cos^{2} \left( \vartheta /2 \right) -
            \left( q-2 \right)  \left( -1+p \right) }{q \left( 1-p \right) 
               \sin^{2} \left( \vartheta /2 \right)   
             }} \right),
              \end{equation}
              provided that it exists. Hence,  the following compact expression for the optimal value of the QFI associated with the weight parameter is obtained:
              \begin{equation}\label{QFIOPT}
                      (F_{\vartheta})_{\text{opt}}=\dfrac{1}{\text{sin}^{2}\vartheta},
                         \end{equation}
  which is clearly independent of $ p$ and $q $. Therefore,  although performing the measurements may vary the acceleration at which the optimal estimation occurs, the optimal value of the QFI, is completely unaffected by the strength of the measurements. Moreover, Eq. (\ref{generalROPT})
  gives us a proper criterion for existence of the QFI optimal value. If the substituted parameters lead to unphysical values for the acceleration parameter $ r $, we conclude that no optimal value is achievable for the QFI in terms of $ r $. Under this situation, i.e., absence of any optimal value for the QFI, we find that if both measurements are performed simultaneously with equal strength (p=q), the  QFI is always lower bounded via $ F^{p=q}_{\vartheta}>1   $, showing enhancement of estimation compared to the case that no measurements are carried out 
  \par
  In the context of applying both  measurements, we generally consider two important regimes, i.e., $ \vartheta>\pi/2 $ and  $ \vartheta<\pi/2 $
  for investigating the optimal behaviour of the QFI in terms of $ p $ or $ q $ when other parameters are constant. Solving equations
   $ \dfrac{\partial F_{\vartheta}}{\partial p}=0 $ and $ \dfrac{\partial F_{\vartheta}}{\partial q}=0 $,  we can obtain the following expressions for $ p_{opt} $ and $ q_{opt} $, respectively:
   
   \begin{equation}\label{POPT}
       p_{opt}={\frac {q\cos \left( 2\,r \right) +2\, \left( q-1 \right)   \csc^{2}
        \left( \vartheta /2 \right)   -3\,q+4}{q\cos \left( 2\,r
        \right) -q+2}}~~~~~~~~~~~~~~~~ \vartheta>\pi/2,
         \end{equation}
         \begin{equation}\label{QOPT}
             q_{opt}={\frac {2\, \left( p-2 \right) \cos \left( \vartheta  \right) -2\,p}{2
             \, \left( p-1 \right) \cos \left( 2\,r \right)   \sin^{2} \left( 
             \vartheta /2 \right)   + \left( p-3 \right) \cos \left( 
             \vartheta  \right) -p-1}} ~~~~\vartheta<\pi/2,
               \end{equation}
 denoting when $ \vartheta>\pi/2~(\vartheta<\pi/2) $, the QFI reveals  optimal behaviour in terms of $ p~(q) $. Figures \ref{POPTfig} and \ref{QOPTfig} illustrate how  these optimal points vary  in terms of other parameters. Especially, Fig. \ref{POPTq} shows  strengthening the postmeasurement, we should increase the strength of the premeasurement for achieving the optimal value of the QFI. Moreover, as seen in Fig. \ref{POPTvar}, larger values of   $ \vartheta $ (the parameter that should be estimated), need larger values of $ p $ for obtaining the optimal value. On the other hand, Fig. \ref{POPTr}
  shows that  more weak premeasurements are required for attaining the optimal QFI when the accelerated observer moves with more larger acceleration. Finally, Fig. \ref{QOPTfig} illustrates the behaviour of $ q_{opt} $ as functions of $ p $, $ \vartheta $ and $ r $. As plotted in 
  Fig. \ref{QOPTp}, strengthening the premeasurement requires increase of the postmeasurement  strength   for achieving the optimal value of the QFI. Besides, contrary to the previous case, smaller values of   $ \vartheta $ need more strong postmeasurements for achieving the optimal estimation
  (see Fig. \ref{QOPTvar}). Moreover, when the  accelerated observer moves with more larger acceleration, we should strengthen the postmeasurement 
   for attaining the optimal QFI (see Fig. \ref{QOPTr}).

  It should be noted that substituting  $ p_{opt} $~($ q_{opt} $) with $ p $~($ q $) in Eq. (\ref{qfivar}), we obtain the optimal value similar to one given in Eq. (\ref{QFIOPT}). Therefore, in both regimes, i.e., $ \vartheta>\pi/2 $ and $ \vartheta<\pi/2 $, achieving the optimal value, we can improve the parameter estimation compared to the scenario in which no measurements are carried out.   
  \subsection{Phase parameter estimation}
                          \begin{figure}[ht!]
         \subfigure[]{\includegraphics[width=8cm]{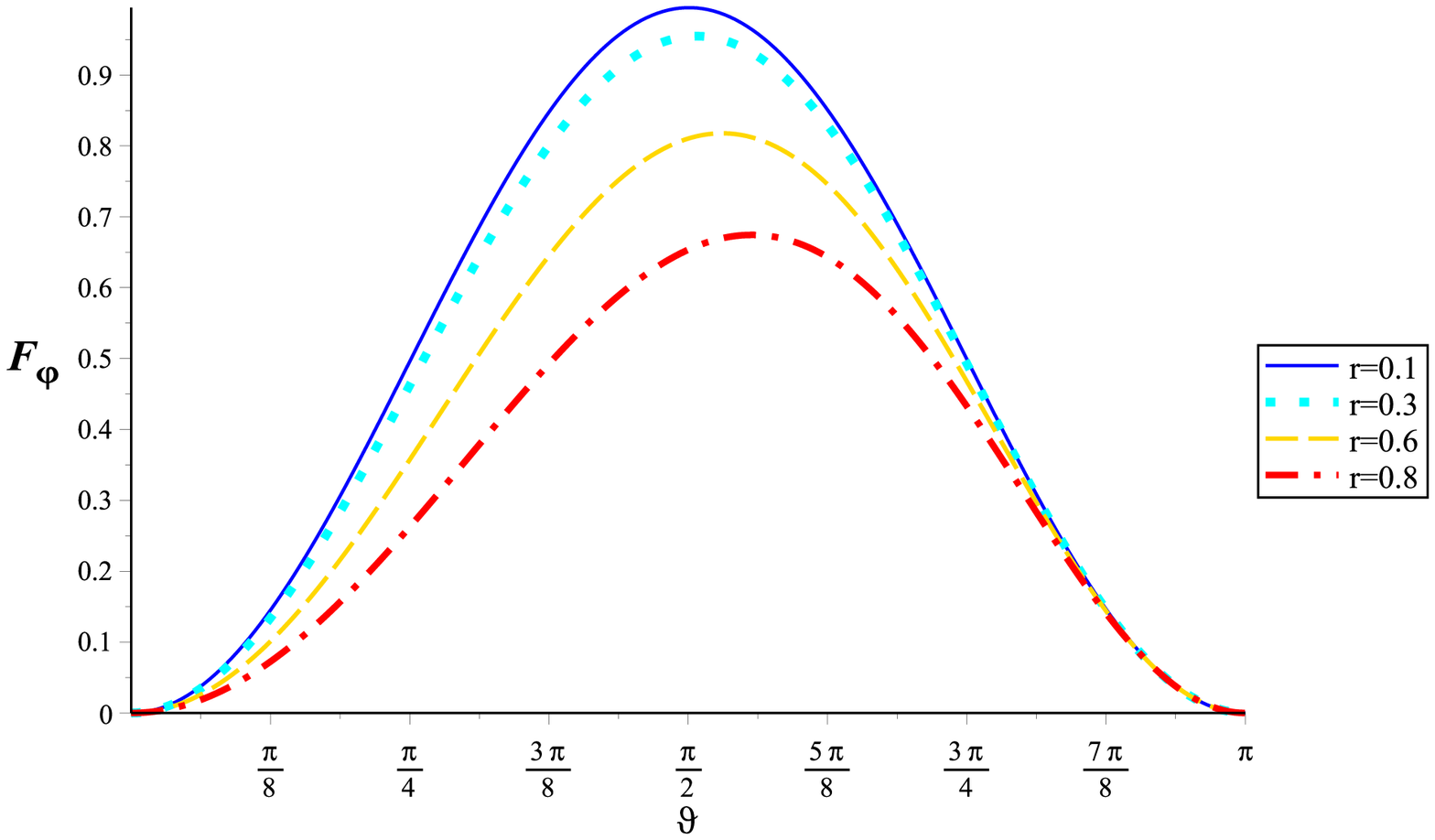}\label{Fphir} }
         \subfigure[]{\includegraphics[width=8cm]{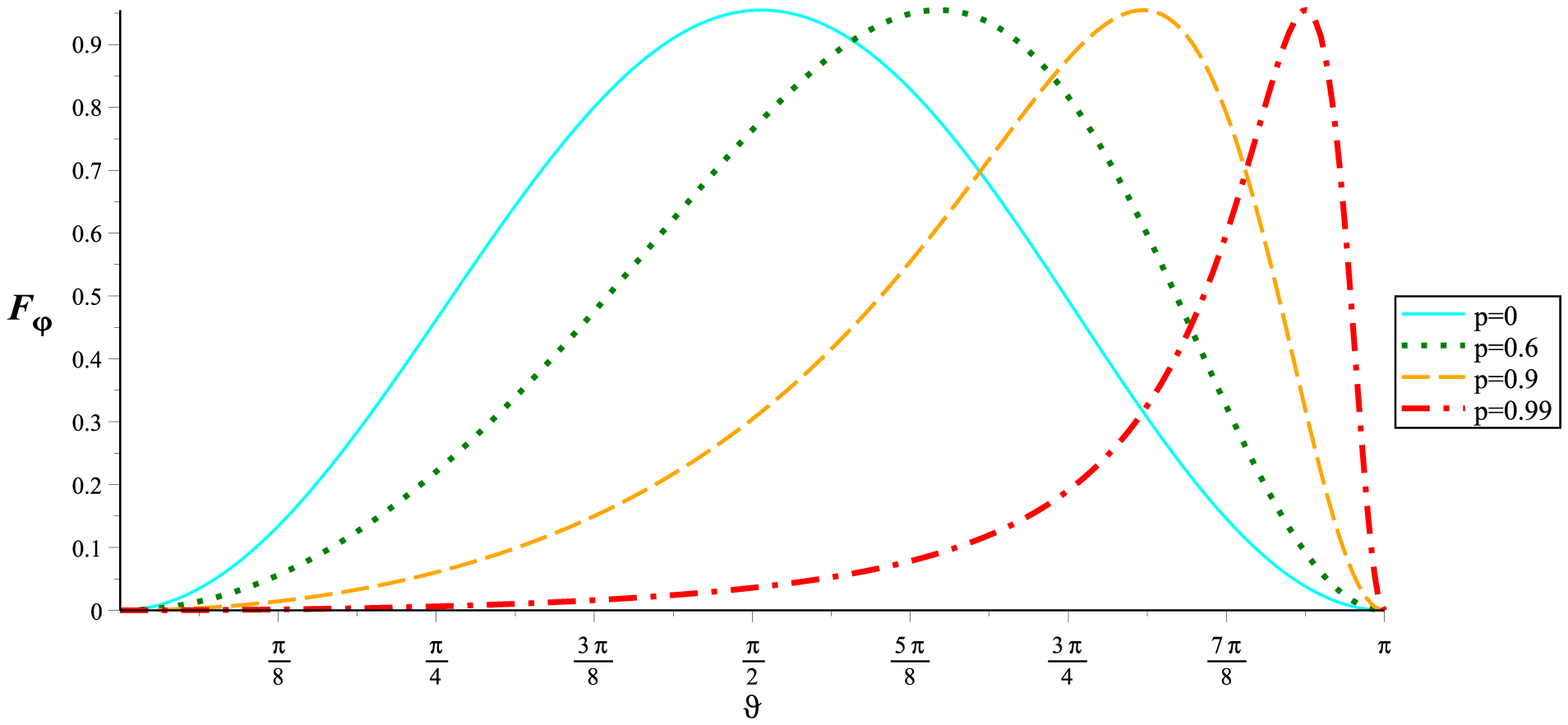}\label{Fphip} }
         \subfigure[]{\includegraphics[width=9cm]{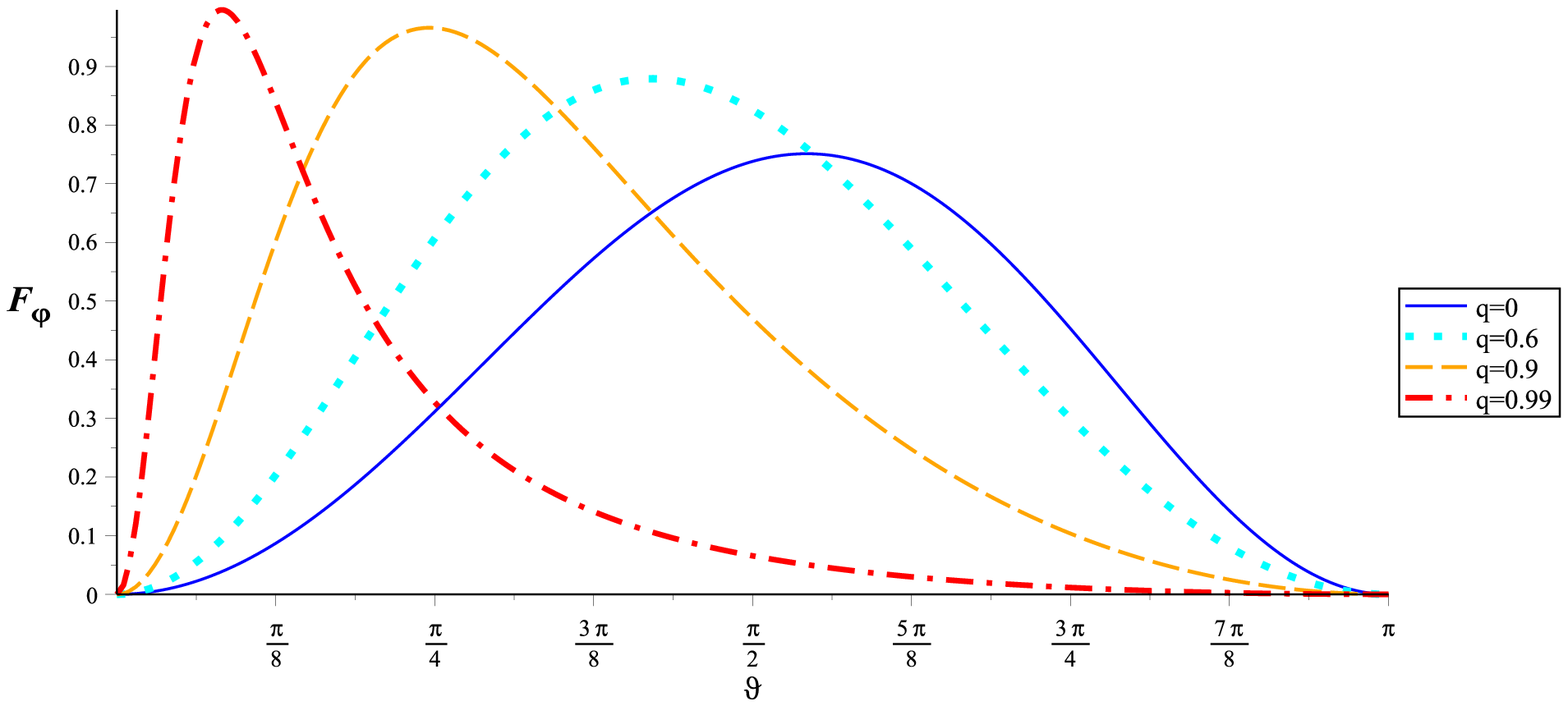}\label{Fphiq} }
             \caption{\small The QFI associated with the phase parameter  as a function of $ \vartheta $; (a) The QFI for $ p=q=0 $ and 
             different values of $ r $. (b) The same quantity for $ r=0.2,~q=0 $, and different values of $ p $. (c) The same quantity for $ r=0.7,~p=0 $, and different values of $ q $}
            
                                              \label{QOPTfig}
                                                \end{figure}
  Estimating phase parameter $ \varphi $ leads to the following expression for the corresponding QFI:
  
   \begin{equation}\label{qfiphi}
  F_{\varphi} ={\frac {-2~\overline{pq}  \cos^{2} \left( r \right)   \bigg( \cos
   \left( 2\,r \right)  \left( \cos \left( \vartheta  \right) -1
   \right) \overline{p}-\overline{p}-2\,\overline{q}+ \left( \overline{p}-2\,\overline{q} \right) \cos \left( \vartheta 
   \right)  \bigg)   \tan^{2} \left( \vartheta /2 \right)   }{ \bigg( 2\,\overline{p}  \cos^{2} \left( r \right)    
  \sin^{2} \left( \vartheta /2 \right)   +2\,\overline{pq} \sin^{2}
   \left( r \right)     \sin^{2} \left( \vartheta /2
   \right)   +\overline{q}+\overline{q}\cos \left( \vartheta  \right)  \bigg) 
   \big( \overline{p}  \cos^{2} \left( r \right)    \tan^{2}
   \left( \vartheta /2 \right)   +\overline{q} \big) ^{2}}}
    \end{equation}
  In order to discuss the optimal behaviour of the QFI, it is assumed that we have no control over the initial state. First,   the QFI variation 
  versus the acceleration parameter when no measurements are made is investigated (see Fig. \ref{Fphir}). It is observed, in the absence of any measurement, when the accelerated observer moves with larger acceleration, the optimal value of the QFI decreases and occurs for  a larger value of the initial parameter $ \vartheta $.
  \par
   Because there is no control over $ \vartheta $, we intend to match the optimal $ \vartheta $ to the predetermined $ \vartheta $. Figures  \ref{Fphip} and Fig. \ref{Fphiq} illustrates how this strategy may be implemented by controlling the strength of the measurements. Figure  \ref{Fphip} shows that increase of the premeasurement strength shifts the optimal point to the right and the same time it does not considerably change the optimal value of the QFI. On the other hand, from  Fig. \ref{Fphiq}, we find that increase of the postmeasurement strength shifts the optimal point to the left and interestingly raises the optimal value of the QFI, leading to enhancement of the phase parameter estimation compared to the case that no measurements are carried out. Overall, using weak measurements, we can control the optimal point at which the QFI is maximized, such that it  coincides with the initial value defined in Eq. (\ref{step1}).
  
  \subsubsection{ Lower bound on QFI with LQU}
  \begin{figure}[ht!]
                                 \subfigure[]{\includegraphics[width=6cm]{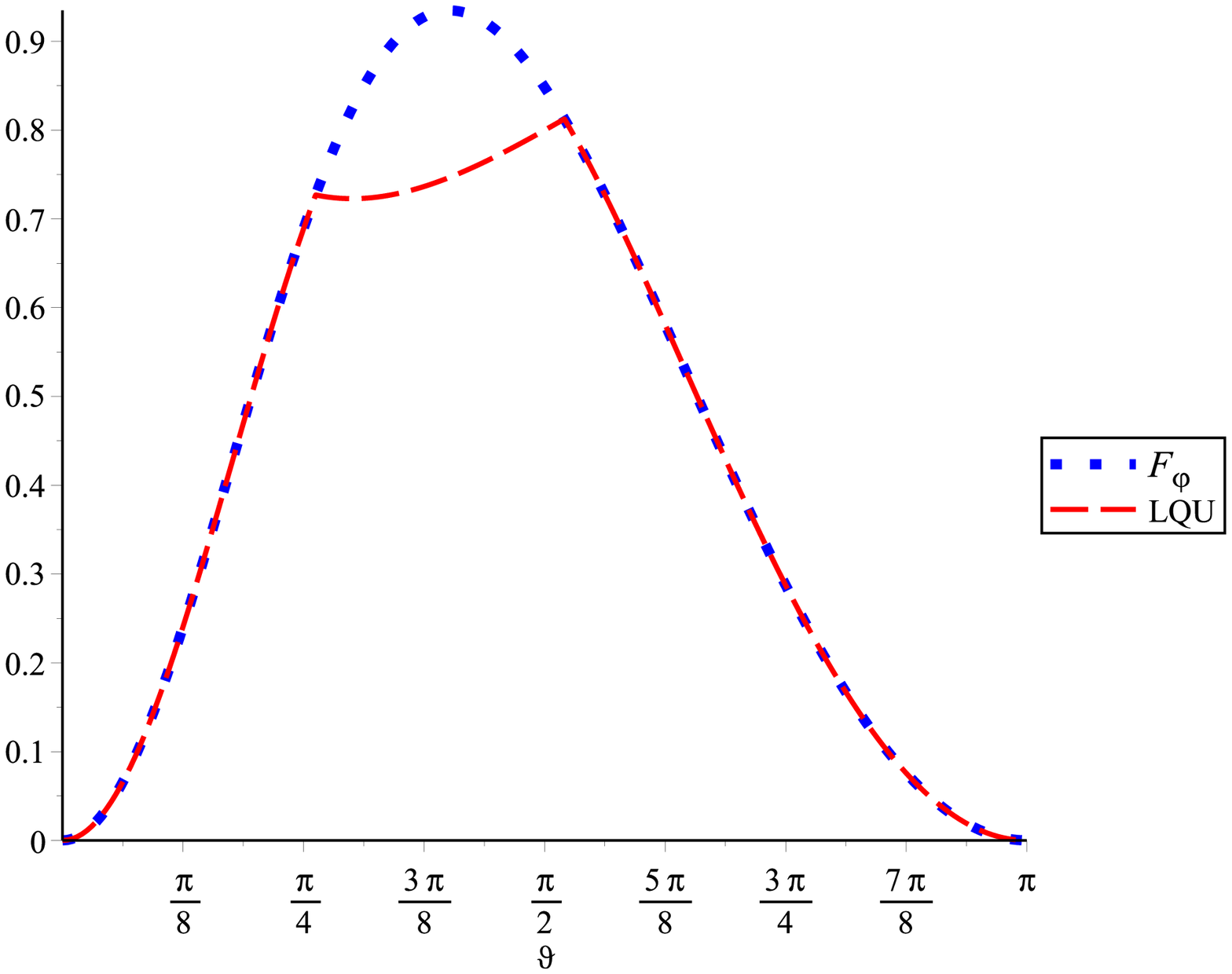}\label{compar1}}
                                 \subfigure[]{\includegraphics[width=6cm]{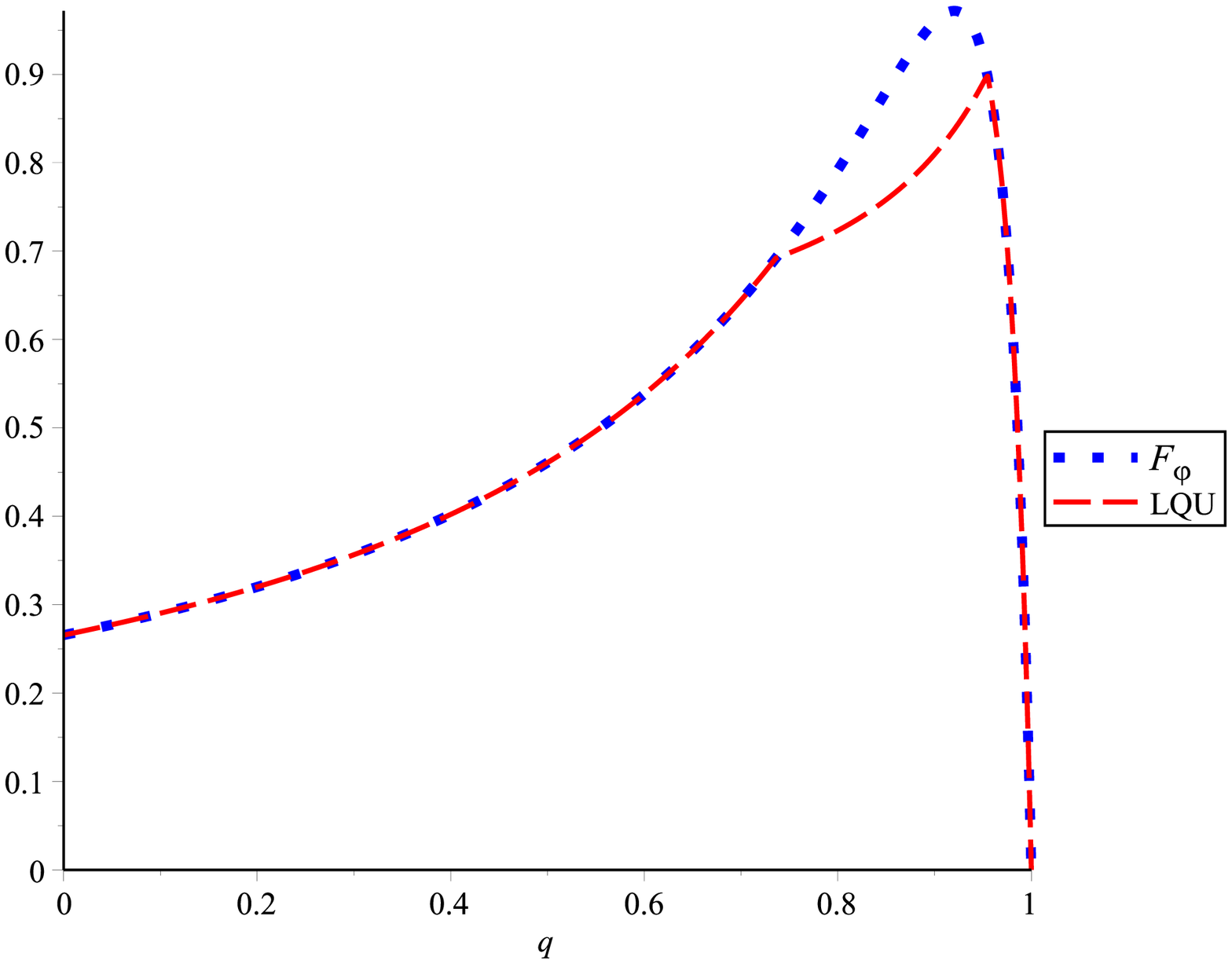}\label{compar2}}
                                 \subfigure[]{\includegraphics[width=6cm]{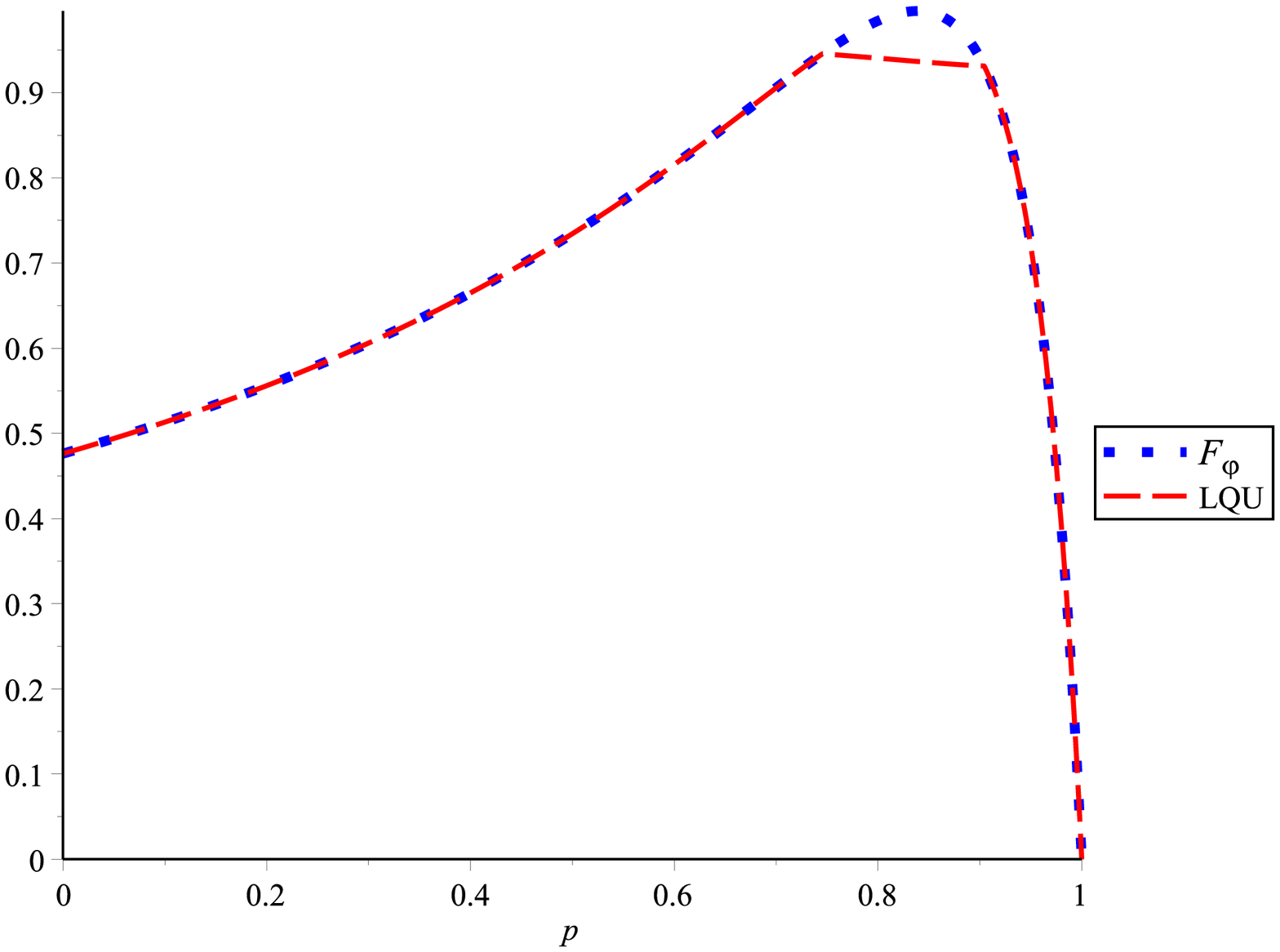}\label{compar3}}
                                  \subfigure[]{\includegraphics[width=6cm]{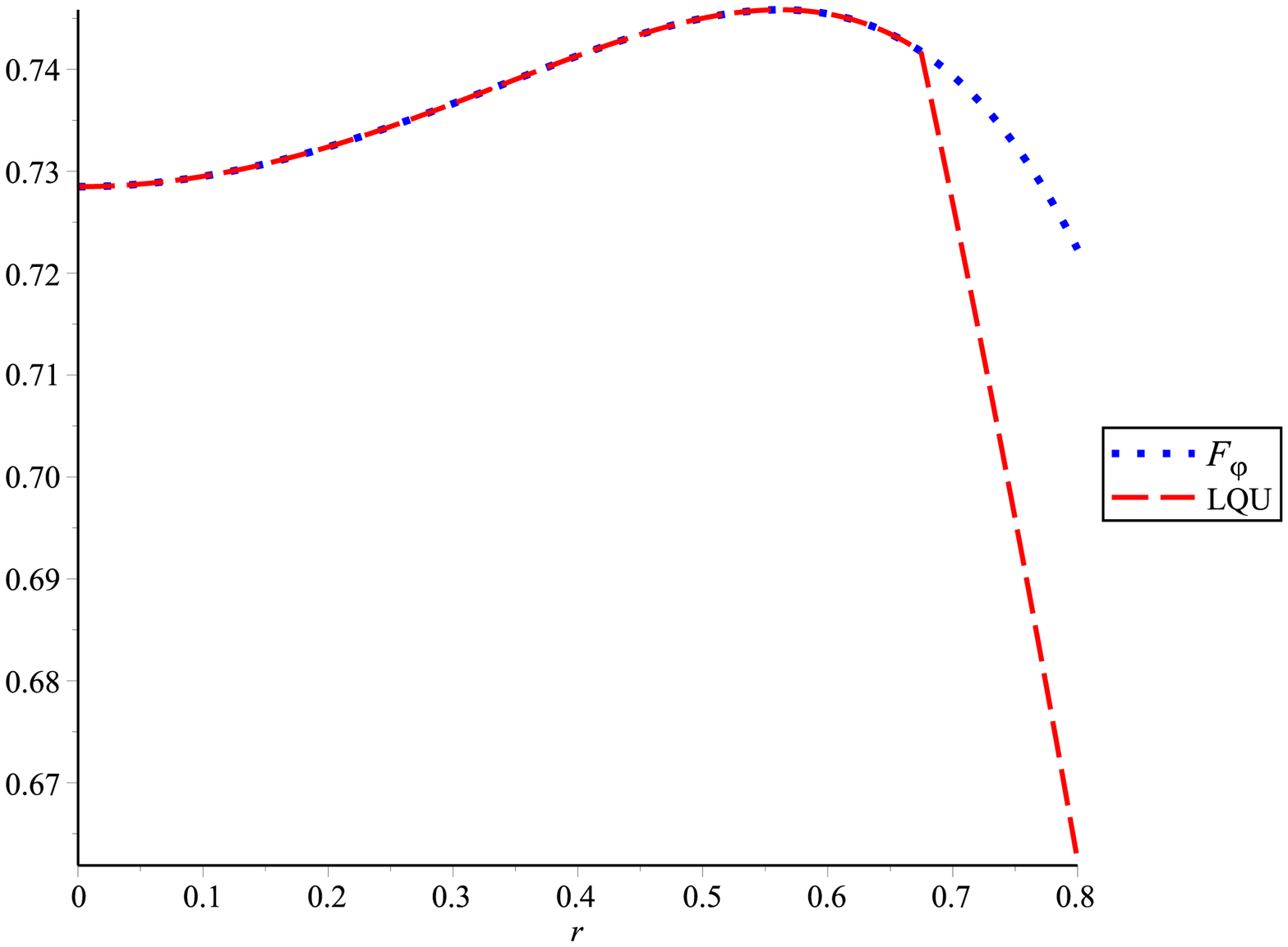}\label{compar4}}
                                     \caption{\small Comparison between   QFI associated with the phase parameter and LQU. (a)
                                     Both quantities as functions of $ \vartheta $ for $ p = 0.4 $, $ r =0 .7 $, and  $ q = 0.8 $.
                                     (b) The same quantities as  functions of $ q $ for $ p = 0.4 $, $ r = 0.7 $, and  $ \vartheta = 0.9 $.
                                     (c) The same quantities as  functions of $ p $ for $ q = 0.2 $, $ r =0 .1 $, and  $ \vartheta = 2.3 $.
                                     (d)The same quantities as  functions of $ r $ for $ q = 0.5 $, $ p = 0 $, and $ \vartheta = 1.8 $.}
                                                                      \label{comparison}
   \end{figure}
The analytical expression for LQU of quantum state (\ref{a16}) is given in Appendix \ref{B}.
 We first briefly review an important discussion from \cite{Girolami2013}.
  Given a (generally mixed) bipartite state $ \rho $ used as a
  probe, subsystem A experiences a unitary transformation so that the bipartite state transforms
  into $ \rho _{\varphi}=U_{\varphi}\rho U^{\dagger}_{\varphi}$, where $U_{\varphi}=\text{exp}^{-i\varphi H_{A}}  $, with $ H_{A} $ a local
  Hamiltonian on A, which we assume to have a nondegenerate spectrum. The initial state (\ref{step1}) may be prepared by the above scenario. 
 Under this condition,  it has been shown that the QFI associated with the phase shift
   majorizes the skew information of the Hamiltonian and therefore the LQU \cite{Girolami2013,Luo2004}, i.e., 
    $\mathcal{U}_{A}(\rho)\leq I(\rho,H_{A})=I(\rho_{\varphi},H_{A})\leq\frac{1}{4}F_{\varphi}$.
   It should be noted  that in this model, the other
   steps of the estimation process are assumed to be noiseless. Besides, in \cite{Shao2018},  it has been shown that for a model, consisting of two qubits interacting with independent non-Markovian environments, when we consider the effects of the environment and the coupling interaction between subsystems, the QFI and  LQU
   do not satisfy the similar relationship generally.
    Now we investigate whether or not  this bound holds for our model including the PMs and the Unruh effect.
    
    \par
     Here, our numerical computation shows that  the
   QFI associated with the phase parameter is bounded from below by the LQU which is  the measure of amount of quantum correlations in the mixed state (\ref{a16}) used for parameter estimation (see Fig. \ref{comparison}), i.e., 
     \begin{equation}\label{inequal}
         \mathcal{U}_{A}(\rho_{A,I})\leq F_{\varphi}.
 \end{equation}
  Especially, if the QFI  reveals no optimal behaviour as a function of one of the parameters shown in Figs. \ref{compar1}-\ref{compar4}   for determined values of other  parameters,  equality $F_{\varphi}= \mathcal{U}_{A}(\rho_{A,I}) $ holds. Moreover, inequality (\ref{inequal})
  refers to the fact that the quantum correlations measured by LQU,  are a sufficient resource to ensure that some information about the phase parameter  can be extracted from the system in the process of estimation. In addition, 
   for probe quantum states with any nonzero amount of discord,
  and for $ M\gg 1 $ repetitions of the estimation process, the optimal
  detection strategy,  asymptotically saturating the quantum Cram\'{e}r-Rao bound, leads to  production of an estimator $ \tilde{\varphi}_{best} $ with
  the following necessarily limited variance:
  
 \begin{equation}\label{VARLQU}
   Var(\tilde{\varphi}_{best})=\dfrac{1}{MF_{\varphi}}\leq \dfrac{1}{M\mathcal{U}_{A}(\rho_{A,I})}     
 \end{equation}
 Therefore, we find that, in the relativistic framework, the quantum correlations measured by the LQU, are a sufficient resource to 
  guarantee an upper bound on the smallest possible variance
  with which the  phase parameter can be measured in the presence of PMs and Unruh effect.

  \subsubsection{ Upper bound on QFI with MSC}
  Using Eq. (\ref{MSCasli}) and density matrix (\ref{a16}), we obtain the following expression for MSC 
   \begin{equation}\label{MYM}
    \Lambda(\rho_{A,I})= {\frac {1}{\sqrt {q+ \left( 1-q \right)  \sec^{2} \left( r \right) 
      }}}
   \end{equation}
 where it is completely unaffected by the premeasurement and initial preparation of the system. Moreover, although the acceleration degrades the MSC, strengthening the postmeasurement can improve it (see Fig. \ref{MSCfig}). In particular, when $ q \rightarrow 1 $, the MSC becomes robust against the Unruh effect.  

 \begin{figure}[ht]
                \includegraphics[width=10cm]{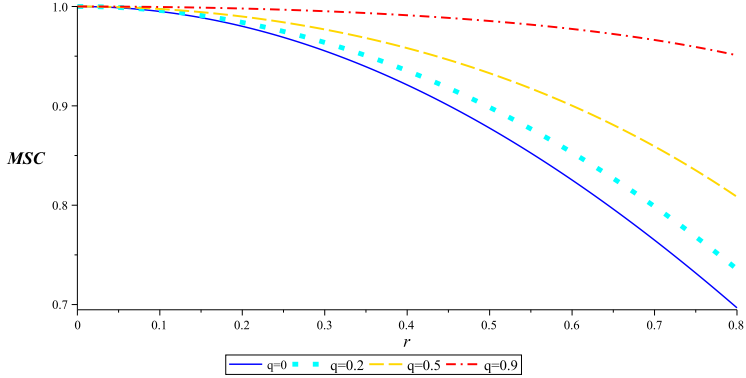}
                \caption{\small    MSC versus $ r $ for different values of $ q $.}
                \label{MSCfig}
                  \end{figure}
                  \begin{figure}[ht!]
                                                   \subfigure[]{\includegraphics[width=6cm]{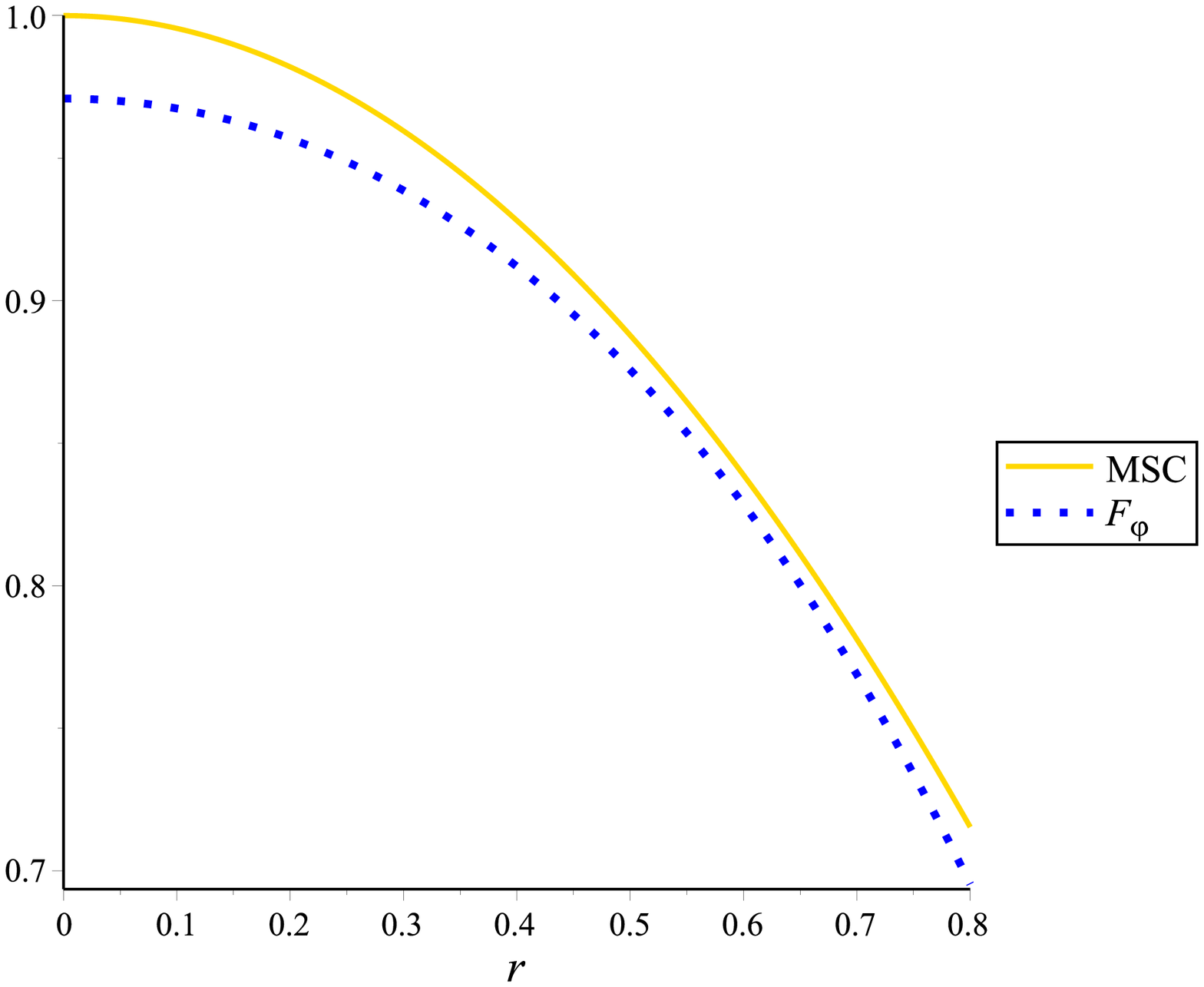}\label{MSCQM1}}
                                                   \subfigure[]{\includegraphics[width=6cm]{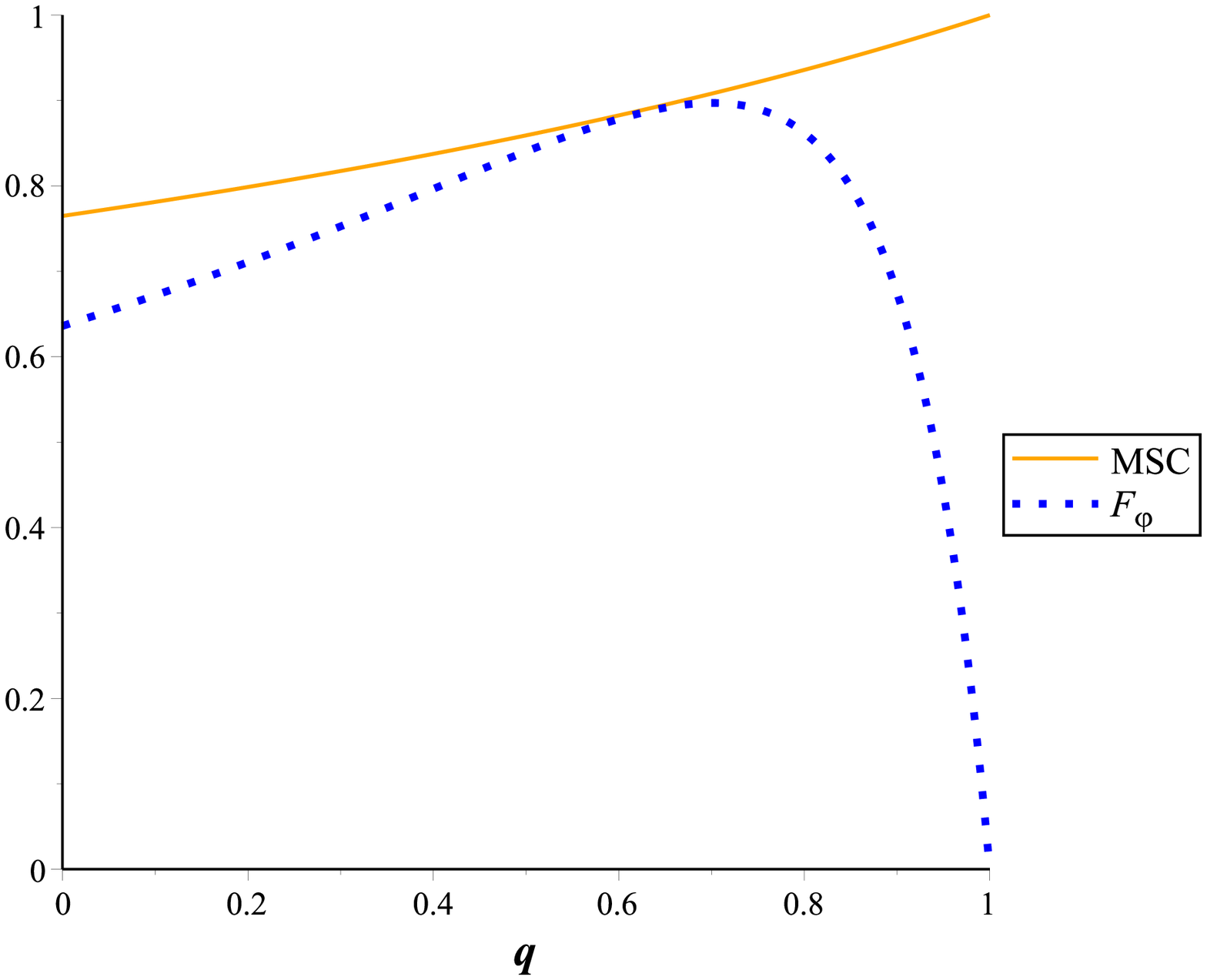}\label{MSCQM2}}
                                                       \caption{\small MSC as an upper bound on the QFI (a)
                                                       Both quantities as functions of $ r$ for $ p = 0.2 $, $ \vartheta  =1.8 $, and  $ q = 0.1 $.
                                                       (b) The same quantities as  functions of $ q $ for $ p = 0.2 $, $ r = 0.7 $, and  $ \vartheta = 1.8 $.}
                                                                                        \label{MSCQM}
                     \end{figure}
 
 Generally, the exact computation of the QFI
 is difficult because it usually needs diagonalization of the
  density matrix. Hence, resorting to upper bounds on the QFI may be beneficial
 both theoretically and practically \cite{Benatti2014,Escher2011}. Now, we reveal an important relationship between the QFI associated with the phase parameter and the MSC. As seen in Fig. \ref{MSCQM}, the MSC determines an  upper bound on the QFI associated with the phase parameter.  
  In particular, this upper bound is approximately saturated near optimal point $ q_{opt} $ at which the QFI is maximized (see Fig. \ref{MSCQM2}).
  \section{Summary and conclusions \label{conclusion}}
  \par To summarize,   we have completely discussed the optimal behaviour of the QFI for  two
   modes of a Dirac field detected by  an inertial observer Alice  and a uniformly accelerated observer
Rob, experiencing the Unruh effect. In particular, we have investigated   the effects of weak measurements, performed before and after accelerating Rob, on the optimal estimation of the weight and phase parameters  of the initial state of the system. 
It was found  that the weak measurements may partially compensate the decreasing effects of the Unruh decoherence on the accuracy of the parameter estimation. 
This enhancement of estimation could  be attributed 
to the probabilistic nature of weak measurements, because in the case  of  projective measurement  ($ p $ or $ q $=1), the QFI jumps to zero and no information can be extracted from the process of quantum estimation. On the other hand,
estimating   weight parameter $ \vartheta $, 
we obtain the QFI trapping  with  value $ F^{p=q=0}_{\vartheta} = 1 $ in the absence of PMs.

 In the case of 
estimating   weight parameter $ \vartheta $, it was interestingly demonstrated that the second PM operates as a quantum key for manifestation of the Unruh effect such that when $ q=0 $, the QFI is unaffected by the Unruh decoherence. Under this situation, the first PM  may guarantees the enhancement of
the estimation provided that $\vartheta > \pi/2 $. Besides, we illustrated that it is possible to design the PMs such that   the optimal estimation  is realized. It was found that achieving the optimal value of the QFI, we can improve the parameter estimation compared to the
scenario in which no measurements are carried out. Moreover, we found that if both measurements are performed simultaneously with equal strength $ (p=q) $, the estimation  is enhanced compared to the case that no measurements are performed. In the case of optimizing
 the phase estimation, nevertheless   we have  no control over  the initial state,
 it was shown that
   the weak measurements may be used to match the optimal $ \vartheta $ to  its predetermined  value.

   Besides, it was found that the QFI associated with the phase parameter is bounded from below by the LQU. More interestingly, our numerical calculation revealed that  the MSC can be utilized to obtain an   upper bound, approximately saturable,  on the QFI in our relativistic scenario. We also obtained a compact formula for the MSC of X-states.

Finally,   It is worth noting that
the PM may be easily applied to different types of qubits such as spin qubits, optical polarization qubits, 
and Josephson junction qubits, etc.  The experimental implementation of PM has been realized  in a photonic architecture \cite{Gillett2010,Lee2011}, and Josephson junction \cite{Katz2008}. Alternatively, the PMs could be described with projective measurements in a larger Hilbert space including an ancilla qubit \cite{Paraoanu2011}. Therefore, the performance of the PM on the target qubit is equivalent to the action of von Neumann projective measurement on the ancilla qubit  previously coupled to it. 
Hence, our scheme  can be realized by current technology 
 for different types of qubits.

\section*{Acknowledgements}

I thank Mauro Paternostro and Rosario  Lo Franco for useful discussions and help. I also wish to acknowledge the financial support of the MSRT of Iran and Jahrom University.

\appendix

\section{Analytical
expression of MSC for X-states}\label{A1}

\par  
The density matrix of a two-qubit X state \cite{Obando2015} shared by  Alice and Rob takes the following
form in the computational basis $ \{|00\rangle,|01\rangle,|10\rangle,|11\rangle\} $

\begin{equation}
\rho_{X}=\left(
\begin{array}{cccc}
\rho_{11} & 0&0&\rho_{14}   \\
0 &\rho_{22} &\rho_{23}&0  \\
0 &\rho_{32} &\rho_{33}&0  \\
\rho_{41} & 0&0&\rho_{44}   \\
\end{array} \right)
\end{equation}
where $ \rho_{23}=\rho_{32}^{*} $, $ \rho_{14}=\rho_{41}^{*} $, and $ \sum\limits_{i=1}^{4}\rho_{ii} =1$. 
Using (\ref{center}), we find  the Rob's steering ellipsoid  is centered at

\begin{equation}\label{Xcenter}
 \textbf{C}_{X}=\left(
 \begin{array}{ccc}
 0   \\
 0   \\
 {\frac {\rho_{{11}}\rho_{{33}}-\rho_{{22}}\rho_{{44}}}{ \left( \rho_{{
11}}+\rho_{{22}} \right)  \left( \rho_{{33}}+\rho_{{44}} \right) }}   \\
 \end{array} \right).
\end{equation}
Moreover, Eq. (\ref{ElipsoidQ}) leads to the matrix

\begin{equation}
Q_{X}=\left(
\begin{array}{ccc}
Q_{11} & Q_{12}&0   \\
Q_{21} &Q_{22} &0  \\
0 &0 &Q_{33} \\
\end{array} \right)
\end{equation}
where
\begin{equation}
Q_{11}={\frac {   \left| \rho_{{14}}+\rho_{{23}} \right|  ^{2}
}{ \left( \rho_{{11}}+\rho_{{22}} \right)  \left( \rho_{{33}}+\rho_{{
44}} \right) }}\nonumber
\end{equation}
\begin{equation}
Q_{12}=Q_{21}={\frac {-2~ \text{Im} \left( \rho_{{14}}\rho_{{32}} \right) }{ \left( \rho_{
{11}}+\rho_{{22}} \right)  \left( \rho_{{33}}+\rho_{{44}} \right) }}\nonumber
\end{equation}
\begin{equation}
Q_{22}={\frac {   \left| \rho_{{14}}-\rho_{{23}} \right|  ^{2}
}{ \left( \rho_{{11}}+\rho_{{22}} \right)  \left( \rho_{{33}}+\rho_{{
44}} \right) }}
\end{equation}
\begin{equation}
Q_{33}={\frac { \bigg( 
 \left( \rho_{{11}}-\rho_{{44}} \right) ^{2}- \left( \rho_{{22}}-\rho_{{33}} \right) ^{2}+1-2\,(\rho_{11}+\rho_{44})
 \bigg) ^{2}}{ \bigg(4~\left( \rho_{{11}}+\rho_{{22}} \right) 
 \left( \rho_{{33}}+\rho_{{44}} \right)\bigg) ^{2}}}\nonumber
\end{equation}
The eigenvalues of matrix $ Q $
are the squares of the ellipsoid semiaxes and its eigenvectors give the orientation of these axes \cite{Miln2015}. Because of block form of $ Q_{X}$, one of its semiaxes  is oriented parallel to  $ \textbf{b} $. In detail, when
$ \rho $ is an X-state, the Bloch vector $ \textbf{b} $ lies along an axis of the Rob's steering ellipsoid, and $ \Lambda $ is the length of the longest of the other two semiaxes \cite{Hu2016}. Following this point,  we find that the analytical
expression of MSC for  X-states is given by

 \begin{equation}\label{MSCasli}
 \Lambda(\rho_{X})={\frac { \left| \rho_{23} \right| + \left| \rho_{14} \right| }{\sqrt {
  \left( \rho_{{11}}+\rho_{{22}} \right)  \left( \rho_{33}+\rho_{44} \right) 
 }}}
 \end{equation}

\section{Expression for LQU}\label{B}
After  tedious calculation, one can obtain the LQU for quantum state (\ref{a16}):

  \begin{equation}
  \mathcal{U}_{A}=1-\text{max}(W_{1},W_{2}),
  \end{equation}
  where $ W_{1} $ and $ W_{2} $ are expressed as
  
   \begin{equation}\label{W1}
  W_{1} =\frac {\sin \left( r \right) \sin \left( \vartheta /2 \right) \sqrt {
      \overline{p}{\overline{q}}^{3} \bigg( -\cos \left( \vartheta  \right)  \big( \overline{p}\cos \left( 2
      \,r \right) +\overline{p}-2\,\overline{q} \big) +\overline{p}\cos \left( 2\,r \right) +\overline{p}+2\,\overline{q}
       \bigg) }}{{\it N}\, \left( \overline{p}  \cos^{2} \left( r \right)  
        \tan^{2} \left( \vartheta /2 \right)   +\overline{q} \right) }
   \end{equation}
  and 
  \begin{equation}\label{W2}
    W_{2} =\dfrac{l_{1}+l_{2}+l_{3}}{{4N}\, \left( \overline{p}  \cos^{2} \left( r \right)  
            \tan^{2} \left( \vartheta /2 \right)   +\overline{q} \right)^{2}}
     \end{equation}
  in which $ l_{i} $'s are given by
  \begin{equation}
     l_{1} ={\overline{q}}^{2} \left( \overline{p}+2\,\overline{q} \right) -\overline{p}\cos \left( 2\,r \right)  \left( \cos
      \left( \vartheta  \right) -1 \right)  \bigg( \overline{q}-\overline{p}  \cos^{2} \left( r
      \right)    \tan^{2} \left( \vartheta /2 \right) 
        \bigg) ^{2},
     \\\nonumber
       \end{equation}
   \begin{equation}
     l_{2} =- \left( \overline{p}-2\,\overline{q} \right) \cos \left( \vartheta  \right)  \bigg( \overline{q}-\overline{p}
        \cos^{2} \left( r \right)     \tan^{2} \left( 
       \vartheta /2 \right)    \bigg) ^{2}+\overline{p} \left( \overline{p}+2\,\overline{q}
        \right)   \cos^{2} \left( r \right)     \tan^{2}
        \left( \vartheta /2 \right)    \left( \overline{p}  \cos^{2}
        \left( r \right)    \tan ^{2} \big( \vartheta /2
        \big)  -2\,\overline{q} \right), 
       \\\nonumber
         \end{equation}
          \begin{equation}
                l_{3} =4\,\overline{p}\overline{q}  \sin^{2} \left( r \right)    \sin^{2} \left( 
                \vartheta /2 \right)   \bigg( \overline{p}  \cos^{2} \left( r
                 \right)    \tan^{2} \left( \vartheta /2 \right) 
                  +\overline{q} \bigg) ^{2}.
                \\\nonumber
                  \end{equation}
\pagebreak

\end{document}